\documentclass[usenatbib, usegraphicx]{mn2e}
\newcommand{\Hebeta}{He$\beta$}
\newcommand{\Lbol}{\ensuremath{L_{\mathrm{bol}}}}
\newcommand{\Lsol}{\ensuremath{L_{\odot}}}
\newcommand{\Lx}{\ensuremath{L_{\mathrm{X}}}}
\newcommand{\Lyalpha}{Ly$\alpha$}
\newcommand{\Lybeta}{Ly$\beta$}
\newcommand{\Lygamma}{Ly$\gamma$}
\newcommand{\Lydelta}{Ly$\delta$}
\newcommand{\fir}{\textit{fir}}
\newcommand{\mH}{\ensuremath{m_{\mathrm{H}}}}
\newcommand{\Mdot}{\ensuremath{\dot{M}}}
\newcommand{\Msol}{\ensuremath{M_{\odot}}}
\newcommand{\Ne}{\ensuremath{n_{\mathrm{e}}}}
\newcommand{\nH}{\ensuremath{n_{\mathrm{H}}}}
\newcommand{\NH}{\ensuremath{N_{\mathrm{H}}}}
\newcommand{\Rsol}{\ensuremath{R_{\odot}}}
\newcommand{\Rstar}{\ensuremath{R_{\ast}}}
\newcommand{\Te}{\ensuremath{T_\mathrm{e}}}
\newcommand{\Tmax}{\ensuremath{T_\mathrm{max}}}

\newcommand{\MdotWR}{\ensuremath{\Mdot_\mathrm{WR}}}
\newcommand{\MdotO}{\ensuremath{\Mdot_\mathrm{O}}}
\newcommand{\vWR}{\ensuremath{v_\mathrm{WR}}}
\newcommand{\vO}{\ensuremath{v_\mathrm{O}}}
\newcommand{\MWR}{\ensuremath{M_\mathrm{WR}}}
\newcommand{\MO}{\ensuremath{M_\mathrm{O}}}

\newcommand{\rchisq}{\ensuremath{\chi^2_\nu}}

\newcommand{\mA}{\ensuremath{\mbox{~m\AA}}}
\newcommand{\angstrom}{\ensuremath{\mbox{~\AA}}}
\newcommand{\cm}{\ensuremath{\mbox{~cm}}}
\newcommand{\km}{\ensuremath{\mbox{~km}}}
\newcommand{\AU}{\ensuremath{\mbox{~A.U.}}}
\newcommand{\pc}{\ensuremath{\mbox{~pc}}}
\newcommand{\s}{\ensuremath{\mbox{~s}}}
\newcommand{\yr}{\ensuremath{\mbox{~yr}}}
\newcommand{\ev}{\ensuremath{\mbox{~eV}}}
\newcommand{\kev}{\ensuremath{\mbox{~keV}}}
\newcommand{\erg}{\ensuremath{\mbox{~erg}}}
\newcommand{\K}{\ensuremath{\mbox{~K}}}
\newcommand{\MK}{\ensuremath{\mbox{~MK}}}

\newcommand{\cc}{\ensuremath{\cm^3}}
\newcommand{\pcc}{\ensuremath{\cm^{-3}}}
\newcommand{\pcmsq}{\ensuremath{\cm^{-2}}}
\newcommand{\ps}{\ensuremath{\s^{-1}}}
\newcommand{\pyr}{\ensuremath{\yr^{-1}}}
\newcommand{\ergps}{\ensuremath{\erg \ps}}
\newcommand{\kmps}{\ensuremath{\km \ps}}
\newcommand{\Msolpy}{\ensuremath{\mbox{ } \Msol \pyr}}

\newcommand{\ArXVII}{Ar\,\textsc{xvii}}
\newcommand{\CII}{C\,\textsc{ii}}
\newcommand{\CIII}{C\,\textsc{iii}}
\newcommand{\CIV}{C\,\textsc{iv}}
\newcommand{\HeI}{He\,\textsc{i}}
\newcommand{\HeII}{He\,\textsc{ii}}
\newcommand{\MgXI}{Mg\,\textsc{xi}}
\newcommand{\MgXII}{Mg\,\textsc{xii}}
\newcommand{\NeII}{Ne\,\textsc{ii}}
\newcommand{\NeIX}{Ne\,\textsc{ix}}
\newcommand{\NeX}{Ne\,\textsc{x}}
\newcommand{\SIV}{S\,\textsc{iv}}
\newcommand{\SXV}{S\,\textsc{xv}}
\newcommand{\SXVI}{S\,\textsc{xvi}}
\newcommand{\SiIII}{Si\,\textsc{iii}}
\newcommand{\SiIV}{Si\,\textsc{iv}}
\newcommand{\SiXIII}{Si\,\textsc{xiii}}
\newcommand{\SiXIV}{Si\,\textsc{xiv}}
\newcommand{\FeXXV}{Fe\,\textsc{xxv}}

\newcommand{\asca}{\textit{ASCA}}
\newcommand{\chandra}{\textit{Chandra}}
\newcommand{\einstein}{\textit{Einstein}}
\newcommand{\rosat}{\textit{ROSAT}}
\newcommand{\xmm}{\textit{XMM-Newton}}

\title[Probing the wind-wind collision in $\gamma^2$~Velorum with high-resolution \chandra\ X-ray spectroscopy]
      {Probing the wind-wind collision in $\gamma^2$~Velorum with high-resolution \chandra\ X-ray spectroscopy:
	evidence for sudden radiative braking and non-equilibrium ionization}
\author[D. B. Henley, I. R. Stevens \& J. M. Pittard]
  {David B. Henley,$^1$\thanks{Email: dbh@star.sr.bham.ac.uk} Ian R. Stevens$^1$ and Julian M. Pittard$^2$  \\
   $^1$School of Physics and Astronomy, University of Birmingham, Edgbaston, Birmingham B15 2TT \\
   $^2$School of Physics and Astronomy, University of Leeds, Woodhouse Lane, Leeds LS2 9JT}

\begin{document}

\maketitle

\begin{abstract}
We present a new analysis of an archived \chandra\ HETGS X-ray spectrum of the WR+O colliding wind binary
$\gamma^2$~Velorum. The spectrum is dominated by emission lines from astrophysically abundant elements:
Ne, Mg, Si, S and Fe. From a combination of broad-band spectral analysis and an analysis of line flux
ratios we infer a wide range of temperatures in the X-ray emitting plasma ($\sim$4--40\MK).
As in the previously published analysis, we find the X-ray emission lines are essentially
unshifted, with a mean FWHM of $1240 \pm 30 \kmps$. Calculations of line profiles based on hydrodynamical
simulations of the wind-wind collision predict lines that are blueshifted by a few hundred\kmps. The lack
of any observed shift in the lines may be evidence of a large shock-cone opening half-angle ($> 85 \degr$),
and we suggest this may be evidence
of sudden radiative braking. From the $R$ and $G$ ratios measured from He-like forbidden-intercombination-resonance
triplets we find evidence that the \MgXI\ emission originates from hotter gas
closer to the O star than the \SiXIII\ emission, which suggests that non-equilibrium ionization may be
present.
\end{abstract}



\section{Introduction}
\label{sec:Introduction}

$\gamma^2$~Velorum (WR 11, HD 68273) is the closest known Wolf-Rayet (WR) star, at a \textit{Hipparcos}-determined distance of
$258^{+41}_{-31} \pc$ \citep*{schaerer97}. As such, it is a key system for increasing
our understanding of X-ray emission from massive early-type stars. $\gamma^2$~Vel is a double-lined
spectroscopic binary of spectral type WC8 + O7.5 \citep{demarco99} whose orbit is well determined, with
a period of $78.53 \pm 0.01$~days, $e = 0.326 \pm 0.01$, $\omega_\mathrm{WR} = 68\degr \pm 4\degr$ \citep{schmutz97} and
$i = 63\degr \pm 8\degr$ \citep{demarco99}.

In a system consisting of a WR star with an early-type companion, a substantial contribution to the X-ray
emission should come from the collision of the stars' dense, highly supersonic winds \citep{cherepashchuk76}.
However, the situation is complicated by the fact that the individual stars may also have substantial intrinsic
X-ray emission from shocks that develop in the winds as a result of line-driven instabilities
\citep*[e.g.][]{owocki88}. An important signature of colliding wind emission is phase-locked variability of
the flux and/or hardness of the X-ray emission. The simplest explanation for the variability of these observable
quantities is the variation in the amount of absorption the X-rays suffer as the system moves through its
orbit. Increasing absorption reduces the flux, but it also increases the hardness because softer X-rays are
more strongly absorbed. In an eccentric binary, there may be other factors which give rise to phase-locked
variability. Near periastron, the intrinsic X-ray luminosity may increase because the winds are
denser when they collide. The emission will also be softer if the winds are not travelling at their terminal
velocities, as the shock speeds will be slower near periastron compared with apastron.

\einstein\ observations of $\gamma^2$~Vel found that it was not unusually bright in X-rays, with
$L_\mathrm{X}/L_\mathrm{bol(O)} = 0.44 \times 10^{-7}$ \citep{pollock87} [cf. the mean
$L_\mathrm{X}/L_\mathrm{bol}$ for single stars in the \einstein\ catalogue of O stars
is $2.5 \times 10^{-7}$ (\citealp{chlebowski89b}; see also \citealp{moffat02})]. There was however a suggestion
that the X-rays from  $\gamma^2$~Vel were softer when the O star is in front of the WR star \citep{pollock87}.

The first convincing evidence of the wind-wind collision in $\gamma^2$~Vel came not from X-ray data, but from
\textit{International Ultraviolet Explorer} (\textit{IUE}) and \textit{Copernicus}
 ultraviolet spectra \citep*{stlouis93}. They found that in broad terms the variability
of the UV line profiles could be explained in terms of selective line eclipses of the O star light by the
WR star wind, with the effect being restricted to resonance and low-excitation transitions of species common
in the WC8 wind (e.g. \CII, \CIII, \CIV, \SiIII, \SiIV\ and \SIV). However, the details of some of the variability
could not be explained by a simple spherically symmetric WR star wind. Instead there was evidence from
resonance lines expected to occur in the winds of both stars (\CIV, \SiIV\ and \SIV) that the collision
between the stars' winds forms a cavity in the WR star wind. At phases when the O star is in front, one is looking
into the cavity and sees the O star wind in absorption. As the O star wind is $\sim$1000\kmps\ faster than the
WR star wind, a high-velocity blue absorption wing appears in the P Cygni profile. At phases when the
WR star is in front, the O star wind travelling towards the observer is prevented from reaching its
terminal velocity by the wind-wind collision, and so the high-velocity wing disappears, masked by the
absorption trough of the WR star wind.

X-ray emission from the wind-wind collision in $\gamma^2$~Vel was first studied in detail with \rosat. From an
analysis of 13 observations, \citet*{willis95} found significant phase-dependent X-ray variability that is
repeatable with binary phase. They found two main components to the emission. When the O star was not in
front of the WR star, the emission in the 0.1--2.5\kev\ range was relatively constant and soft ($kT \sim 0.19\ \kev$)
with $\Lx \sim 2.5 \times 10^{31} \ergps$. When the O star was in front they found the X-ray emission was
enhanced by the addition of a harder component, with $kT \ga 1$--2\kev\ and $\Lx \ga 10^{32} \ergps$.
They attributed this harder component to emission from the wind-wind collision, observed through the cavity
in the WR star wind formed by the O star wind. This was supported by hydrodynamical modelling of the wind-wind
collision, which found that at \rosat\ energies the emission from the wind-wind collision is only observable
at phases when the O star is in front. Using the winds' terminal velocities in the hydrodynamical simulations
overpredicted the luminosity in the \rosat\ band (0.1--2.5\kev) by about an order of magnitude. Better
agreement with the observations was obtained by assuming the O star wind collided at less than its
terminal velocity, which is expected as the wind-wind collision region is likely to be close to the O star
(as its wind is much weaker than that of the WR star).

The hydrodynamical modelling in \citet{willis95} predicted copious X-ray emission at energies greater than
2.5\kev, inaccessible by \rosat, but within the range of \asca. \citet{stevens96} obtained two \asca\
spectra taken near periastron (at phases $\Phi = 0.978$ and $\Phi = 0.078$, where $\Phi = 0$ corresponds
to periastron).	These were compared with a set of synthetic X-ray spectra, generated from hydrodynamical
simulations with a range of different wind parameters (namely the mass-loss rates and terminal velocities
of the two winds). By fitting their synthetic spectra to their observed spectra, \citet{stevens96} were not
only able to confirm the conclusion of \citet{willis95} that $\gamma^2$~Vel is a colliding wind system, but
were also able to put constraints on some of the stars' wind parameters. Most notably, they found
the mass-loss rate of the W-R star to be $\sim 3 \times 10^{-5} \Msolpy$. This is a factor
of three lower than the mass-loss rate derived from radio observations \citep*[$8.8 \times 10^{-5} \Msolpy$][]{barlow88}.
\citet{stevens96} suggest this may be because radio observations tend to overestimate mass-loss rates if the wind is
inhomogeneous.

\citet{rauw00} analysed an additional \asca\ spectrum of $\gamma^2$ Vel, this time near apastron
($\Phi = 0.570$), and also reanalysed the data in \citet{stevens96}. Having
taken into account background sources, \citet{rauw00} effectively fitted the hard variable
emission with a single temperature model. The bulk of the variability can be attributed to a changing
column density towards the source. However, the temperature and luminosity also vary.
The $\Phi = 0.978$ spectrum has a lower temperature ($kT \sim 1.9 \kev$, using non-solar abundances)
than the $\Phi = 0.570$ spectrum ($kT \sim 2.7 \kev$). This could be because near apastron the winds
are moving more quickly when they collide, resulting in a larger post-shock temperature. The softest
(and intrinsically most luminous) spectrum is seen at $\Phi = 0.078$ ($kT \sim 1.3 \kev$). If the soft
X-rays are from the wind-wind collision, then at this phase they are being seen through the cavity in the WR
star wind caused by the wind collision (whose position has been deflected by the Coriolis force).
Alternatively, it could indicate that the intrinsic emission from the O star wind is making a significant
contribution. In either case, at the other two phases the soft emission is absorbed by the wind of the WR star.

The unprecedented spectral resolution offered by the \chandra\ Low- and High-Energy Transmission Grating
Spectrometers (LETGS and HETGS) and the \xmm\ Reflection Grating Spectrometer ($E / \Delta E \sim 100$--1000)
give us the opportunity to measure X-ray line shifts and widths to accuracies down to a few hundred\kmps.
This enables us to probe in detail the dynamics of the X-ray-emitting plasma in colliding
wind binaries, giving new insights into the structure of the wind-wind interaction regions in such systems.
                                                  
We present a new analysis of an archived \chandra\ HETGS observation of $\gamma^2$~Vel. These data have already been
analysed by \citet{skinner01}. They found that the X-ray emission lines were broadened
[full width at half-maximum $(\mathrm{FWHM}) \sim 1000 \kmps$]
but unshifted from their rest wavelengths. Furthermore, their results imply
that the \NeIX\ line emission originates tens of stellar radii from the O star, well away from the central
wind-wind collision region near the line of centres. Our analysis includes a larger number of emission lines
than published by \citet{skinner01}. We also discuss in more detail the interpretation of our results in terms
of a colliding wind picture.
The details of the observation and the data reduction are described in Section~\ref{sec:Observation}. The broadband
spectral properties are described in Section~\ref{sec:Broadband}, while the results of fitting to the individual
emission lines are described in Section~\ref{sec:LineFitting}. In Section~\ref{sec:Modelling} we describe
our attempts to model the emission line profiles using the model of \citet*{henley03}. Our results are discussed
in Section~\ref{sec:Discussion} and summarized in Section~\ref{sec:Summary}. Throughout this paper, the quoted errors
are $1 \sigma$ confidence ranges.


\section{Observation and data reduction}
\label{sec:Observation}

$\gamma^2$~Vel was observed by the \chandra\ HETGS between 2000 March 15
09:20:04 UT and 2000 March 16 04:08:12 UT, giving 64\,848 s of useful exposure
time. The observation covered orbital phases $\Phi = 0.080$--0.090, where
$\Phi = 0$ corresponds to periastron passage, using the ephemeris of \citet{schmutz97}:
\begin{equation}
	\mbox{JD(periastron)} = 2\,450\,120.5 + 78.53E
\label{eq:Ephemeris}
\end{equation}
where $E$ is the number of orbits since the periastron passage on JD\,2\,450\,120.5.
Mid-observation was 6.7 days after periastron, and about 4 days after the passage of the O star
in front of the WR star ($\Phi = 0.03$). The separation of the stars during
the observation was 0.92--0.94\AU\ (assuming $\MWR = 9.5 \Msol$ and $\MO = 30 \Msol$;
\citealp{demarco99,demarco00}).

\begin{figure*}
\centering
\includegraphics[width=16cm]{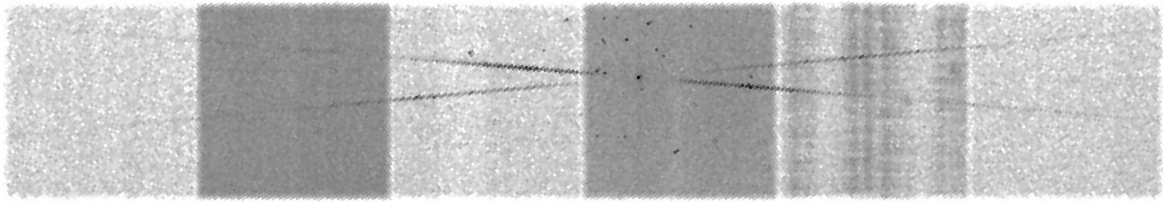}
\caption{The \chandra\ ACIS-S detector showing the dispersed photons of the HETGS spectrum of $\gamma^2$~Vel.
One arm of the X is from the HEG, the other is from the MEG. For the purposes of this figure, the data have
been binned up by a factor of 16 to exaggerate the grating arms, and the image has been rotated so the ACIS-S
detector is horizontal.}
\label{fig:RawImage}
\end{figure*}

The dispersed photons from the \chandra\ HETGS grating assembly form a shallow X in the focal plane
(see Fig.~\ref{fig:RawImage}).
One `arm' of the X is from the High-Energy Grating (HEG), the
other is from the Medium-Energy Grating (MEG). The HEG has higher spectral resolution than the MEG:
$\Delta \lambda = 12 \mA$ FWHM over the range 1.2--15\angstrom\ for the HEG, versus
$\Delta \lambda = 23 \mA$ FWHM over the range 2.5--31\angstrom\ for the MEG (\chandra\ Proposers'
Observatory Guide\footnote{http://cxc.harvard.edu/proposer/POG/index.html}, Section~8.1). On the other hand, the MEG
gives a higher signal-to-noise ratio in the first-order spectra of $\gamma^2$~Vel.

The data were reduced from the Level 1 events file using \textsc{ciao} v3.1,
following threads available from the \chandra\ website\footnote{http://cxc.harvard.edu/ciao/threads/gspec.html}.
This was done in preference to using the spectrum file obtained from the \chandra\
archive because it meant a more up-to-date version of CALDB (v2.27) could be applied to the data.
Similarly, response files (ARFs and RMFs) appropriate for this observation were generated following
\chandra\ threads, in preference to using `off-the-shelf' response files.

There are a total of 7850 counts in the non-background-subtracted first-order HEG spectrum and 16003 counts
in the non-background-subtracted first-order MEG spectrum. For the analysis described here, the data were
not background subtracted, nor was the background spectrum modelled separately.
For each of the two HETGS gratings, the background counts are extracted from two
regions either side of and 4.5 times as wide as the source extraction regions (which are 4.8~arcsec wide in
the cross-dispersion direction). There is therefore a possibility
that photons from bright spectral lines could spread out into the background extraction regions (because of the
point-spread function of the telescope). Indeed, there is some evidence that the brighter emission lines in
the source spectrum are also in the background spectrum. This means that background subtraction could adversely
affect the measured results. Furthermore, the Cash statistic, which was used for the analysis of emission lines described in
Section~\ref{sec:LineFitting} cannot be used with background-subtracted data. However, the count rates in the background
extraction regions are less than 1 per cent of the rates in the corresponding source extraction regions (taking
into account the fact that the background counts come from a detector area 9 times larger than the source counts).


\section{Broad-band spectral properties}
\label{sec:Broadband}

\begin{figure*}
\centering
\includegraphics[width=18cm]{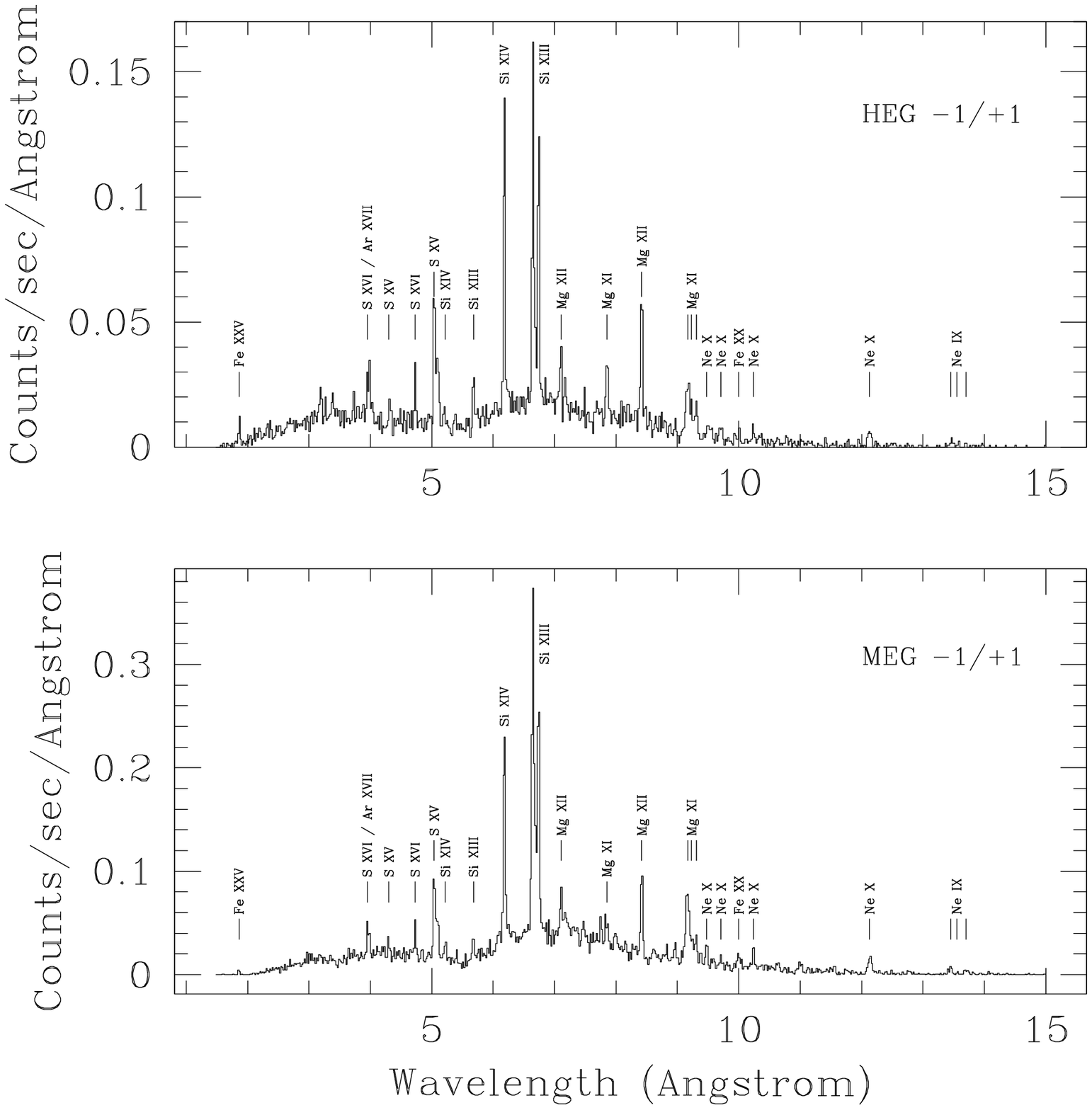}
\caption{The first-order HEG  and MEG  spectra of $\gamma^2$ Velorum. The $+1$ and $-1$ orders have
been co-added, and the spectra binned up to 0.02\angstrom.}
\label{fig:BroadbandSpectrum}
\end{figure*}

The HEG and MEG spectra of $\gamma^2$~Vel are shown in Fig.~\ref{fig:BroadbandSpectrum}. For illustrative purposes,
the $+1$ and $-1$ orders have been co-added, and the spectra have been binned up to 0.02\angstrom. The spectrum
of $\gamma^2$~Vel is dominated by strong emission lines from S ($\lambda \approx 4$--5\angstrom), 
Si ($\lambda \approx 5$--7\angstrom) and Mg ($\lambda \approx 7$--9\angstrom),
with weaker lines from Ne ($\lambda \approx 9$--14\angstrom) and Fe (e.g. $\lambda \approx 2 \angstrom$ and
$\lambda \approx 10 \angstrom$). In this section we describe the analysis of the global properties of the 
spectrum, while in the following section we describe our analysis of the individual emission lines.

The broad-band spectral analysis was carried out using \textsc{xspec}\footnote{http://heasarc.gsfc.nasa.gov/docs/xanadu/xspec/}
v11.3.1. For the purposes of this analysis, the $+1$ and $-1$ orders of each spectrum were co-added, and the data were
binned so there were at least 20 counts per bin. This meant that the $\chi^2$ statistic could be used, and the goodness of
various spectral models could be assessed. The spectral models were fitted to the HEG and MEG spectra simultaneously, using
all the available data.

The ARFs used in the broad-band spectral fitting were generated by co-adding the ARFs for the relevant $+1$ and $-1$ orders. A similar
procedure cannot be followed for the RMFs, and so one must use the RMF for either the $+1$ or $-1$ order
for the whole co-added spectrum. Experimentation showed that the combination of HEG and MEG RMFs used did
not have a significant effect on the results. However, it was noted that the value of $\chi^2$ obtained when
using the HEG~$-1$ RMF and the MEG~$+1$ RMF was always lower than when using any of the other three possible combinations.
The differences in $\chi^2$ were always small, but it may be connected to the fact
that the HEG~$-1$ order and the MEG~$+1$ order contain the larger number of counts in their respective spectra
[3951 counts (HEG~$-1$) versus 3899 counts (HEG~$+1$) and 8543 counts (MEG~$+1$) versus 7460 counts (MEG~$-1$)].
The results described hereafter were obtained using this pair of RMFs.

\subsection{2$T$ models}
\label{subsec:2Tapec}

A visual inspection of the \chandra\ HETGS spectrum of $\gamma^2$~Vel indicates there is a wide
range of temperatures present in the X-ray-emitting plasma, as evidenced by emission lines
from a wide range of ions [from \NeIX, whose emission peaks at $\Tmax \approx 4 \MK$ ($k\Tmax = 0.34 \kev$),
to \SXVI, whose emission peaks at $\Tmax \approx 25 \MK$ ($k\Tmax = 2.2 \kev$)]. Hydrodynamical
simulations of the wind-wind collision also predict a wide range of temperatures (see Section~\ref{sec:Modelling}).
One would therefore not expect a 2$T$ thermal plasma model to provide a good fit to the spectrum; at best it
can only characterize the emission, rather than accurately describe in detail the temperature structure of the
X-ray-emitting gas. Furthermore, because of the wealth of detail in a grating spectrum
(as opposed to a CCD spectrum), a plasma emission
model may have difficulty accurately fitting all the emission lines. The problems that arise could be due to several
factors: Doppler shifting of lines, line broadening (on top of thermal Doppler broadening and instrumental broadening),
non-solar abundances, and inaccurate atomic physics parameters (such as transition rates and lab wavelengths)
in the plasma emission model. Nevertheless, with these caveats in mind, fitting a 2$T$ model to the spectrum of
$\gamma^2$~Vel can give some insight into the general properties of the X-ray-emitting plasma. For this purpose,
we used the \textsc{xspec} \texttt{apec} model (Astrophysical Plasma Emission
Code\footnote{http://cxc.harvard.edu/atomdb/sources\_apec.html}; \citealp{smith00,smith01}).
To model the absorption by the stellar winds we used the \textsc{xspec} \texttt{wabs} model
(which uses cross-sections from \citealp{morrison83}).

We expect a significant portion of the
X-ray emission to originate from the wind-wind interaction region, which is an approximately conical region around
the O star whose apex points along the line of centres towards the WR star. The hot shocked gas streams along this
cone away from the apex. At the time of the \chandra\ observation, the O star was approximately in front of the WR
star, and hence the opening of the shock cone was towards the observer. There was therefore hot gas streaming along
the shock cone towards the observer, and so we would expect to observe blueshifted emission lines. Because the
opening of the shock cone was not directly towards the observer, there would in fact have been a range of
line-of-sight velocities (distributed azimuthally around the shock cone), and so we would also expect to
observe broadened emission lines. This simple picture of the X-ray emission from the wind-wind collision is
developed further in Section~\ref{subsec:Geometry}.

The amount of Doppler shifting and Doppler broadening of the X-ray emission lines  can be determined to a certain
degree from the broad-band spectrum (though this is done in more detail by investigating the individual lines; see
Section~\ref{sec:LineFitting}). Line Doppler shifts can be modelled by thawing the \texttt{apec} model's redshift
parameter $z$. Note that this parameter measures the line-of-sight velocity of the X-ray emitting plasma, not
the systemic line-of-sight velocity (as the shocked gas is moving with respect to the stars).
Line broadening can be modelled by using the \textsc{xspec} \texttt{gsmooth} model, which convolves the whole
spectrum with a Gaussian. The standard deviation $\sigma$ of this Gaussian can vary as a power-law with photon energy $E$:
\begin{equation}
        \sigma(E) = \sigma_6 \left( \frac{E}{6\kev} \right)^\alpha
\label{eq:gsmooth}
\end{equation}
where $\sigma_6$ is the standard deviation at 6\kev. However, in practice it was found that thawing the index $\alpha$
did not significantly improve the fit, so it was frozen at 1 (which gives a constant line width expressed as a velocity).
For the fitting, one \texttt{gsmooth} component was used
for the whole spectrum, and also the redshifts of the two \texttt{apec} models were constrained to be equal
(so there is one redshift characterizing the whole spectrum).

We used non-solar abundances in our fitting, using the \textsc{xspec} \texttt{vapec} model. This model calculates
abundances relative to hydrogen, which is essentially absent in WR winds. We overcame this by setting a large
helium abundance [$N(\mathrm{He})/N(\mathrm{H}) = 10^6$, where $N$(X) denotes the number abundance of element X],
and then measuring abundances relative to helium. The abundances of C, N and O were fixed at
$N$(C)/$N$(He) = 0.14 \citep{morris99,schmutz99}, $N$(N)/$N$(He) = $1.0 \times 10^{-4}$ \citep{lloyd99} and
$N$(O)/$N$(He) = 0.028 [from $N$(C)/$N$(O) = 5; \citealp{demarco00}]. Furthermore, as a starting point we
used $N$(Ne)/$N$(He) = $3.5 \times 10^{-3}$ and $N$(S)/$N$(He) = $6 \times 10^{-5}$ \citep{dessart00},
and we fixed the relative abundances of all elements heavier than neon at their solar values \citep{anders89}.
From this starting point, we allowed the abundances of Ne, Mg, Si, S and Fe to vary.

\begin{table*}
\caption{Results of fitting the \chandra\ HETGS spectrum of $\gamma^2$~Vel with an absorbed two-temperature thermal
plasma model [models A and B: \texttt{wabs$*$gsmooth$*$(vapec+vapec)}] and an absorbed differential emission
measure thermal plasma model [model C: \texttt{wabs$*$gsmooth$*$c6pvmkl} (see Section~\ref{subsec:DEM})].
Model A has $\sigma_6$ [see equation~(\ref{eq:gsmooth})] and the redshift frozen at 0; in models B and C they are free
parameters. The emission measures are expressed in terms of the helium number density $n_\mathrm{He}$
($\mathrm{EM} = \int \Ne n_\mathrm{He} \mathrm{d}V$), because of the lack of hydrogen in the WR star wind.
The quoted abundances are number ratios, relative to the solar number ratios \citep{anders89}.
The X-ray luminosities are for the 0.4--10\kev\ energy range.}
\centering
\begin{tabular}{lccc}
\hline
Model & A & B & C \\
\hline
\NH\ ($10^{22} \pcmsq$)                         & $1.92 \pm 0.04$
						& $2.08 \pm 0.03$
						& $2.24^{+0.07}_{-0.04}$ \\
$kT_1$ (keV)                                    & $0.773^{+0.012}_{-0.014}$
						& $0.763^{+0.010}_{-0.009}$
						& -- \\
EM$_1$ ($10^{54} \pcc$)				& $2.29 \pm 0.08$
						& $2.28^{+0.03}_{-0.15}$
						& -- \\
$kT_2$ (keV)                                    & $2.19^{+0.08}_{-0.05}$
						& $2.15^{+0.04}_{-0.05}$
						& -- \\
EM$_2$ ($10^{54} \pcc$)				& $1.07 \pm 0.05$
						& $0.96^{+0.07}_{-0.03}$
						& -- \\
$\sigma_6$ (eV)                                 & 0 (frozen)
						& $10.6^{+0.2}_{-0.4}$
						& $9.3^{+0.3}_{-0.4}$ \\
Redshift $z$ (km~s$^{-1}$)                      & 0 (frozen)
						& $-18.3^{+0.9}_{-1.4}$
						& $-160^{+10}_{-14}$  \\
(Ne/He) / (Ne/He)$_\odot$			& $0.98 \pm 0.12$
						& $1.81^{+0.19}_{-0.18}$
						& $1.53^{+0.17}_{-0.16}$ \\
(Mg/He) / (Mg/He)$_\odot$			& $0.53 \pm 0.03$
						& $1.09^{+0.05}_{-0.04}$
						& $1.13^{+0.04}_{-0.06}$ \\
(Si/He) / (Si/He)$_\odot$			& $0.85^{+0.02}_{-0.03}$
						& $1.34 \pm 0.03$
						& $1.24^{+0.03}_{-0.05}$ \\
(S/He) / (S/He)$_\odot$				& $1.17 \pm 0.07$
						& $1.66^{+0.07}_{-0.08}$
						& $1.50 \pm 0.07$ \\
(Fe/He) / (Fe/He)$_\odot$			& $0.55 \pm 0.06$
						& $0.99^{+0.06}_{-0.05}$
						& $0.68^{+0.06}_{-0.05}$ \\
$\chi^2_\nu$ (d.o.f.)                           & 2.81 (919)
						& 1.78 (917)
						& 1.95 (914) \\
$L_\mathrm{X}^\mathrm{abs}$ ($10^{32} \ergps$)  & 1.0
						& 1.1
						& 1.1 \\
$L_\mathrm{X}^\mathrm{int}$ ($10^{32} \ergps$)  & 6.7
						& 8.4
						& -- \\
\hline
\end{tabular} 
\label{tab:Broadband}
\end{table*}

The results of fitting the \chandra\ HETGS spectrum of $\gamma^2$~Vel with an absorbed two-temperature thermal
plasma model with non-solar abundances [\texttt{wabs$*$gsmooth$*$(vapec+vapec)}] are shown in
Table~\ref{tab:Broadband}. Model A has $\sigma_6$ and $z$ frozen at 0, whereas in model B they are both free parameters.
The data and the best-fitting models (concentrating on the S and Si line region) are shown in 
Figs.~\ref{fig:BroadbandFitA} (model A) and \ref{fig:BroadbandFitB} (model B). Thawing $\sigma_6$ and $z$
significantly improves $\chi^2$, although the fit is still formally unacceptable (the reduced $\chi^2$ is
1.78 for 917 degrees of freedom). Figs.~\ref{fig:BroadbandFitA} and \ref{fig:BroadbandFitB} indicate that the
poor fits are mainly due to the emission lines, rather than the continuum. Nevertheless, the derived
parameters can give us an overview of the general properties of the X-ray-emitting plasma. The 2 temperatures
derived from the fit ($\approx$9 and 25\MK) are sensible given the range of peak emission temperatures of the lines
observed in the spectrum. The value of
$\sigma_6$ corresponds to a FWHM of $1250^{+20}_{-50} \kmps$. The small measured line shift
($z = -18\kmps$) implies the lines are not systematically shifted from their rest wavelengths. The values of $\sigma_6$ and $z$
can be compared with the results of fitting to individual emission lines (see Section~\ref{subsec:Comparison}).
Thawing $\sigma_6$ and $z$ does not affect the temperatures and emission measures of the two components. Also, while
it seems to significantly affect the column density \NH, this is probably not the case because the errors in
Table~\ref{tab:Broadband} are likely to be underestimated (given the poorness of the fits).

While the abundances for a given model in Table~\ref{tab:Broadband} seem well constrained, a comparison between models
indicates that this is in fact not the case. Furthermore, some measurements are hampered by the fact that
for some elements only a few lines are available to constrain the abundance. For example, the neon abundance is
really only being determined from the \NeX\ \Lyalpha\ and \Lybeta\ lines, and this will lead to a temperature-abundance
degeneracy. Indeed, uncertainties in the temperature distribution will affect all the abundance determinations.
Unfortunately, this therefore means that the abundances in Table~\ref{tab:Broadband} are rather unreliable.

\begin{figure*}
\centering
\includegraphics[width=0.7\linewidth]{fig3a.eps}
\includegraphics[width=0.7\linewidth]{fig3b.eps}
\caption{The binned HEG (top) and MEG (bottom) spectra from $\gamma^2$~Vel plotted with the best-fitting absorbed
2$T$ \texttt{apec} model, with the redshift $z$ and $\sigma_6$ frozen at 0 (model A in Table~\ref{tab:Broadband}).
The residuals are the direct differences between the data and the model (in normalized counts~s$^{-1}$~\AA$^{-1}$).
}
\label{fig:BroadbandFitA}
\end{figure*}

\begin{figure*}
\centering
\includegraphics[width=0.7\linewidth]{fig4a.eps}
\includegraphics[width=0.7\linewidth]{fig4b.eps}
\caption{As Fig.~\ref{fig:BroadbandFitA}, but with the redshift $z$ and $\sigma_6$
free to vary (model B in Table~\ref{tab:Broadband}).}
\label{fig:BroadbandFitB}
\end{figure*}

Note that the X-ray luminosities in Table~\ref{tab:Broadband}
have been derived using the \textit{Hipparcos}-determined distance of $258^{+41}_{-31} \pc$
\citep{schaerer97}. This result has recently been thrown into doubt by the serendipitous discovery of an association
of low-mass pre-main sequence (PMS) stars in the direction of $\gamma^2$~Vel \citep{pozzo00}. \citet{pozzo00} argue
that these PMS stars are at approximately the same distance and age as $\gamma^2$~Vel, placing $\gamma^2$~Vel
within the Vela OB2 association at a distance of 360--490\pc, in good agreement with older distance estimates
\citep[e.g. 460\pc;][]{conti72}. If this larger distance is correct, the X-ray 
luminosities in Table~\ref{tab:Broadband} should be increased by a factor of $\sim$3.

The absorbing column \NH\ derived from the fit corresponds
to a visual extinction $A_v \approx 9$, using the empirical relation of \citet{gorenstein75}. In contrast, the
visual extinction derived from optical studies is $A_v = 0.00$--0.12 \citep{vanderhucht01}. This implies that there is
extra absorption local to the source, in addition to the interstellar absorption. Given the orientation of the
orbit and the phase of the observation, the wind-wind collision zone was observed through the wind of the O star.
The value of \NH\ is consistent with the column density through the wind of the O star to a point in the
wind-wind collision zone that lies on the line of centres, which may reasonably be assumed to be a characteristic
column density for the whole emission region. This can be shown by considering the optical depth
$\tau_\nu$ through a spherically symmetric wind expanding at constant velocity $v_\infty$ to a point at
($r$,$\theta$,$\phi$) (in spherical coordinates centred on the star, with the $z$-axis along the line of sight), which
is given by \citep{ignace01a}
\begin{equation}
        \tau_\nu = \tau_\star \frac{R_\star}{r} \frac{\theta}{\sin \theta}
\label{eq:tau_nu1}
\end{equation}
where $R_\star$ is the stellar radius and $\tau_\star = \kappa_\nu \Mdot / 4 \pi v_\infty R_\star$ is a characteristic wind
optical depth \citep{owocki01}, where \Mdot\ is the stellar mass-loss rate and $\kappa_\nu$ is the opacity.
The optical depth is related to the column density \NH\ by
\begin{equation}
        \tau_\nu = \int \kappa_\nu \rho \mathrm{d}r \approx \int \kappa_\nu \mH \nH \mathrm{d}r = \kappa_\nu \mH \NH
\label{eq:tau_nu2}
\end{equation}
where $\rho$ is the mass density, \nH\ is the number density of hydrogen and \mH\ is the mass of a hydrogen atom.
Thus, by combining equations~(\ref{eq:tau_nu1}) and (\ref{eq:tau_nu2}), we obtain
\begin{equation}
        \NH = \frac{\Mdot}{4 \pi \mH v_\infty r} \frac{\theta}{\sin \theta}
\label{eq:NH}
\end{equation}
The angle $\theta$ [that is, the polar angle (in this coordinate system) of a point on the line of centres]
is related to the orbital phase angle $\Psi$ ($\Psi = 0\degr$ corresponding to the
O star being in front) and the orbital inclination $i$ by
\begin{equation}
	\cos \theta = - \cos \Psi \sin i
\label{eq:theta}
\end{equation}
[See also equation~(4) in \citet{henley03}; note that the $\theta$ used in that equation is the supplement of
the $\theta$ in equations~(\ref{eq:tau_nu1}) and (\ref{eq:NH}).]
Using the orbital elements of \citet{schmutz97} ($e = 0.326$, $\omega_\mathrm{WR} = 68\degr$) and
the phase of the \chandra\ observation ($\Phi = 0.085$ at mid-observation; see Section~\ref{sec:Observation}),
we obtain $\Psi = 36\degr$. Combining this with $i = 63\degr$ \citep{demarco99} gives $\theta = 136\degr$.
Using the separation ($\approx$0.93\AU) and the wind momentum ratio ($\MdotWR \vWR / \MdotO \vO \approx 33$; \citealp{demarco00}),
the distance along the line of centres from the O star to the wind-wind collision zone is $r \approx 2 \times 10^{12} \cm$.
The mass-loss rate and wind velocity are more difficult to determine, and the situation is complicated by the fact that
the wind will not be at its terminal velocity at the wind collision. One consequence of this is that the column density
will be larger than that given by equation~(\ref{eq:NH}), because the inner regions of the wind are denser than is assumed by
a constant-expansion model. Assuming $\Mdot = 0.1$--$1 \times 10^{-6} \Msolpy$ and $v_\infty = 2000 \kmps$ (which are sensible
values for a late O star), we find $\NH \approx 0.3$--$3 \times 10^{22} \pcmsq$, which is consistent with the
values derived from the spectral fitting. However, it should be noted that the \texttt{wabs} absorption model in \textsc{xspec}
assumes neutral absorbing material. This will overestimate the opacity of the partially ionized stellar wind, and so
underestimate \NH. This discrepancy will be worst at lower energies ($\la 1 \kev$; see fig.~4 in \citealp{cohen96}).

If we invert the above argument, and specify our measured value of \NH\ in equation~(\ref{eq:NH}),
we obtain $\MdotO \sim 8 \times 10^{-7} \Msolpy$. However, this value may not be very reliable, given the approximations
used to obtain it.

\subsection{DEM modelling}
\label{subsec:DEM}

As has already been stated, a $2T$ \texttt{apec} model can only characterize the spectrum, rather than accurately
describing the temperature structure of the X-ray-emitting plasma. One would expect to obtain a more detailed
picture of the temperature structure by using a differential emission measure (DEM) model.

\citet{skinner01} fitted the first-order MEG spectrum of $\gamma^2$~Vel with an absorbed differential emission
measure (DEM) model, and found a strong emission measure peak near $\sim$9--10\MK.
We attempted to reproduce this, using a \texttt{wabs$*$gsmooth$*$c6pvmkl} model in \textsc{xspec}.
This model is based on the \texttt{mekal} thermal plasma model (unfortunately there is no DEM model
in \textsc{xspec} based on \texttt{apec}). The DEM is expressed as an exponential of a polynomial
\begin{equation}
	\mathrm{DEM}(T) \propto \exp \left[ \sum^6_{k=1} a_k P_k (T) \right]
\label{eq:DEM}
\end{equation}
where $P_k$ is a Chebyshev polynomial of order $k$, and the coefficients $a_k$ are free to vary during the fit.
As in the previous section, the abundances of Ne, Mg, Si, S and Fe were free parameters, as were $\sigma_6$
and $z$. Again, we fitted the HEG and MEG spectra simultaneously.

\begin{figure}
\centering
\includegraphics[width=8cm]{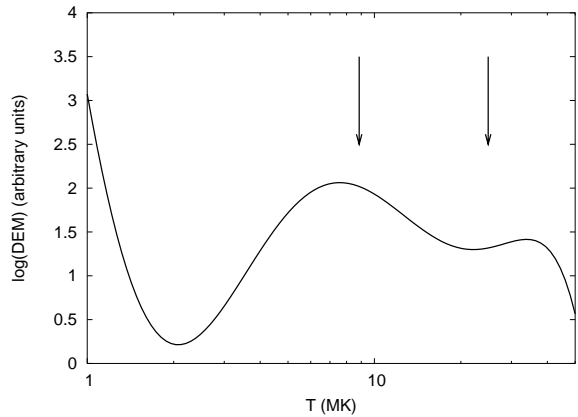}
\caption{Our best fit differential emission measure (DEM) model for $\gamma^2$~Vel.
The arrows (at 8.85\MK\ and 24.9\MK)
indicate the best fit temperatures from our 2$T$ \texttt{apec} model fitting (Table~\ref{tab:Broadband}, model B).
The rise in the DEM below $\sim$2\MK\ is probably an artefact of the assumed functional form of the DEM
[equation~(\ref{eq:DEM})].}
\label{fig:DEM}
\end{figure}

Our best-fitting DEM is shown in Fig.~\ref{fig:DEM}, and the resulting spectral parameters are in
Table~\ref{tab:Broadband} (model C). The DEM \texttt{mekal} model actually gives a worse fit than
the 2$T$ \texttt{apec} model. This is most likely due to the fact that \texttt{mekal} is an older code,
and so may have inaccurate wavelengths, transition rates, etc. compared with the more up-to-date
\texttt{apec} model. This means the derived spectral parameters should be treated with caution (in
particular the redshift $z$, which is significantly different from that obtained in
Section~\ref{subsec:2Tapec}).

Our DEM is broadly peaked at $\sim$7--8\MK, compared with the strong peak at $\sim$10\MK\ found
by \citet{skinner01}. To estimate the uncertainty on the shape of the DEM, we have calculated the
errors on the coefficients $a_k$ using \textsc{xspec}, and then used these in a Monte-Carlo
simulation to generate a set of 1000 DEMs. While the small peak at $\sim$35\MK\ does not seem
to be very significant, the drops in the DEM below 7\MK\ and above 40\MK\ do seem to be real.
The DEM therefore indicates a wide range of plasma temperatures from a few\MK\
to $\sim$40\MK, as expected from the range of emission lines seen in the spectrum.
Without knowing more about \citeauthor{skinner01}'s~\citeyearpar{skinner01} DEM, we cannot say
whether or not our DEM is consistent with theirs.

An alternative method of calculating the DEM (with uncertainties) would be to use the method described
by \citet{wojdowski04}, which involves fitting the spectrum with a number of components, each
of which is the entire X-ray emission line spectrum for a single ion. The DEM is thus pieced
together from the emission measures of these individual components (each of which corresponds
to a different temperature range). However, such an additional analysis of the DEM is beyond
the scope of this paper.

Rather surprisingly, our DEM also shows a pronounced rise
below $\sim$2\MK. However, as there is very little emission above $\sim$14\angstrom, the cool end of the
DEM (below $\sim$4\MK, the peak emission temperature of \NeIX) will be very poorly constrained. This rise
in the DEM is thus almost certainly just an artefact of the assumed functional form of the DEM
[equation~(\ref{eq:DEM})].
An unfortunate consequence of the rise in the DEM at low temperatures is that the
intrinsic (unabsorbed) flux is over-estimated by several orders of magnitude at low energies.
For this reason we have not quoted a value for $L_\mathrm{X}^\mathrm{int}$ for this model
in Table~\ref{tab:Broadband}. This excessive intrinsic low-energy flux probably explains why
the column density derived from the DEM model is somewhat higher than that derived from the 2$T$ model
(see Table~\ref{tab:Broadband}).

\section{Emission Line Properties}
\label{sec:LineFitting}

Unlike the broad-band spectral analysis, the analysis of the individual lines in the \chandra\ spectrum of $\gamma^2$~Vel
was carried out using unbinned, non-co-added spectra, so no spectral information was lost. Because some bins
contained low number of counts, the Cash statistic \citep{cash79} was used instead of $\chi^2$. This analysis was carried
out using \textsc{sherpa} (as distributed with \textsc{ciao} v3.1).

The analysis of the X-ray spectrum concentrated on the emission lines from H- and He-like ions of Ne, Mg, Si and S. For all
four of these elements the H-like \Lyalpha\ lines and He-like \textit{fir} (forbidden-intercombination-resonance) triplets
were detected. \Lybeta\ lines were also analysed for \NeX, \MgXII\ and \SiXIV, but not \SXVI, as its \Lybeta\ line is
blended with the \ArXVII\ \textit{fir} triplet near 3.95\angstrom. The \NeX\ Lyman series extends to include \Lygamma\
and \Lydelta. \Hebeta\ lines
($1\mathrm{s}3\mathrm{p}~^1\mathrm{P}_1 \rightarrow 1\mathrm{s}^2~^1\mathrm{S}_0$) were also detected from \SiXIII\ and \MgXI.

Each emission line or multiplet was analysed individually over a narrow range of wavelengths (typically 0.2--0.5\angstrom).
The fitted model consisted of a continuum component plus Gaussian component(s) to model the line emission. 
Although colliding wind binaries are expected to exhibit a wide range of line profile shapes \citep{henley03},
Gaussians give good fits to the emission lines in $\gamma^2$~Vel (see Section~\ref{subsec:FitResults}),
and thus provide a good way of quantifying the line shifts and widths. It was found that
the choice of continuum model (constant or power-law) did not significantly affect the best-fit parameters of line emission
component. As a result the simpler constant model was chosen in preference. The nature of the line emission
component used depended on the nature of the line being investigated, as discussed below.

Lyman lines are in fact closely spaced doublets due to the spin-orbit splitting of the upper level. 
The splitting is $\approx$5\mA\ for \Lyalpha\ and $\approx$1\mA\ for \Lybeta.
This separation is too small to be resolved by the \chandra\ gratings, and so a good fit can be obtained
with a single Gaussian. However, it is not obvious what should be used as the rest wavelength of the line
if one wishes to measure a Doppler shift. The Lyman lines were therefore
fitted with two Gaussians whose parameters were tied together. The Doppler shifts of the two components
(expressed as velocities) were constrained to be equal, as were their widths. The relative intensities of the
two components were fixed at the theoretical values from \textsc{atomdb} v1.1.0, the shorter-wavelength
component being approximately twice as bright as the longer-wavelength component.

In contrast to the Lyman lines, the components of the He-like \textit{fir} triplets are resolvable by the \chandra\ gratings.
These lines were fitted with three Gaussians whose parameters were tied together (the intercombination line is in fact
a closely spaced doublet, but using four Gaussians did not significantly alter the results). The Doppler shifts
and widths were linked as for the Lyman lines. For this purpose, the laboratory wavelength of the brighter,
longer-wavelength intercombination line was used. The amplitudes of the Gaussians were all free to vary.
The \Hebeta\ lines are singlets, and so these were fitted with single Gaussians.

\subsection{Fits to individual lines}
\label{subsec:FitResults}

The fit results are shown in Table~\ref{tab:LineFitting}. 
All these results were obtained by fitting the line model to all four spectra (HEG~$-1$ and $+1$, MEG~$-1$ and $+1$)
simultaneously, except for the \SXVI\ \Lyalpha\ line (the results were obtained just from the MEG data, as a sensible
fit could not be obtained if the HEG data were included) and the \SiXIV\ \Lyalpha\ line (the MEG~$-1$ data were excluded
from the fit; see below). For each individual spectrum the appropriate ARF and RMF was used.

\begin{table*}
\caption{The measured wavelengths ($\lambda_\mathrm{obs}$), widths ($\Delta \lambda$) and fluxes of the X-ray emission
lines in the \chandra\ HETGS spectrum of $\gamma^2$~Vel. All results were obtained by fitting to the unbinned, non-co-added
HEG and MEG spectra simultaneously (except where indicated).
Parameters without quoted errors were tied to other fit parameters,
rather than being free. Laboratory wavelengths ($\lambda_\mathrm{lab}$) and temperatures of maximum emission (\Tmax)
are from \textsc{atomdb} v1.1.0.}
\centering
\begin{tabular}{llllllll}
\hline
Ion	& Line		& $k\Tmax$	& $\lambda_\mathrm{lab}$ & $\lambda_\mathrm{obs}$ & $\Delta \lambda$ (FWHM) & $\Delta \lambda$ (FWHM)	& Observed flux \\
	&		& (keV)		& (\AA)			 & (\AA)		  & (\AA)		    & (km~s$^{-1}$)		& ($10^{-5}$ photons\pcmsq\ps) \\
\hline
\SXVI$^\mathrm{a}$
  	& \Lyalpha$_1$	& 2.165		& 4.7274                 & $4.7289 \pm 0.0020$    & $0.0115 \pm 0.0068$     & $730 \pm 430$		& $1.92 \pm 0.36$ \\
	& \Lyalpha$_2$	& 2.165		& 4.7328		 & 4.7343		  & 0.0115		    & 730			& 0.92 \\
\SXV   	& He $r$	& 1.366		& 5.0387                 & $5.0382 \pm 0.0010$    & $0.0203 \pm 0.0020$	    & $1210 \pm 120$		& $8.60 \pm 0.69$ \\
	& He $i$	& 1.085		& 5.0665		 & 5.0660		  & 0.0205		    & 1210			& $2.71 \pm 0.51$ \\
	& He $f$	& 1.366		& 5.1015		 & 5.1010		  & 0.0206		    & 1210			& $4.15 \pm 0.50$ \\
\SiXIV 	& \Lybeta$_1$	& 1.366		& 5.2168                 & $5.2168 \pm 0.0036$    & $0.026  \pm 0.010$	    & $1510 \pm 570$		& $1.28 \pm 0.38$ \\
	& \Lybeta$_2$	& 1.366		& 5.2180		 & 5.2179		  & 0.026		    & 1510			& 0.61 \\
\SiXIII	& \Hebeta	& 0.862		& 5.6805                 & $5.6800 \pm 0.0018$    & $0.0199 \pm 0.0047$	    & $1050 \pm 250$		& $2.45 \pm 0.40$ \\
\SiXIV$^\mathrm{b}$
 	& \Lyalpha$_1$	& 1.366		& 6.1804                 & $6.1795 \pm 0.0006$    & $0.0237 \pm 0.0015$	    & $1150 \pm 73$		& $5.94 \pm 0.25$ \\
	& \Lyalpha$_2$	& 1.366		& 6.1858		 & 6.1849		  & 0.0237		    & 1150			& 2.84 \\
\SiXIII	& He $r$	& 0.862		& 6.6479                 & $6.6459 \pm 0.0004$    & $0.0287 \pm 0.0010$	    & $1295 \pm 45$		& $12.23 \pm 0.42$ \\
	& He $i$	& 0.862		& 6.6882		 & 6.6862		  & 0.0289		    & 1295			& $3.13 \pm 0.28$ \\
	& He $f$	& 0.862		& 6.7403		 & 6.7383		  & 0.0291		    & 1295			& $7.98 \pm 0.32$ \\
\MgXII 	& \Lybeta$_1$	& 0.862		& 7.1058                 & $7.1049 \pm 0.0019$    & $0.0247 \pm 0.0077$	    & $1040 \pm 230$		& $0.94 \pm 0.14$ \\
	& \Lybeta$_2$	& 0.862		& 7.1069		 & 7.1060		  & 0.0247		    & 1040			& 0.45 \\
\MgXI  	& \Hebeta	& 0.544		& 7.8503                 & $7.8521 \pm 0.0035$    & $0.0407 \pm 0.0075$	    & $1550 \pm 290$		& $1.11 \pm 0.22$ \\
\MgXII 	& \Lyalpha$_1$	& 0.862		& 8.4192                 & $8.4169 \pm 0.0010$    & $0.0291 \pm 0.0028$     & $1036 \pm 89$		& $2.97 \pm 0.21$ \\
	& \Lyalpha$_2$	& 0.862		& 8.4246		 & 8.4223		  & 0.0291		    & 1036			& 1.42 \\
\MgXI  	& He $r$	& 0.544		& 9.1687                 & $9.1636 \pm 0.0020$    & $0.0501 \pm 0.0042$     & $1640 \pm 140$		& $3.63 \pm 0.29$ \\
	& He $i$	& 0.544		& 9.2282		 & 9.2261		  & 0.0505                  & 1640			& $1.23 \pm 0.22$ \\
	& He $f$	& 0.544		& 9.3143		 & 9.3091		  & 0.0509		    & 1640			& $1.20 \pm 0.21$ \\
\NeX   	& \Lydelta$_1$	& 0.544		& 9.4807		 & $9.4810 \pm 0.0033$	  & $0.0263 \pm 0.0081$     & $830 \pm 260$		& $0.45 \pm 0.10$ \\
	& \Lydelta$_2$	& 0.544		& 9.4809 		 & 9.4812		  & 0.0263		    & 830			& 0.22 \\
\NeX   	& \Lygamma$_1$	& 0.544		& 9.7080		 & $9.7095 \pm 0.0050$	  & $0.0414 \pm 0.0099$	    & $1280 \pm 310$		& $0.41 \pm 0.11$ \\
	& \Lygamma$_2$	& 0.544		& 9.7085		 & 9.7100		  & 0.0414		    & 1280			& 0.20 \\
\NeX   	& \Lybeta$_1$	& 0.544		& 10.2385                & $10.2393\pm 0.0029$    & $0.0319 \pm 0.0080$     & $930 \pm 230$		& $0.70 \pm 0.13$ \\
	& \Lybeta$_2$	& 0.544		& 10.2396		 & 10.2404		  & 0.0319		    & 930			& 0.34 \\
\NeX   	& \Lyalpha$_1$	& 0.544		& 12.1321                & $12.1331\pm 0.0032$    & $0.0594 \pm 0.0075$	    & $1470 \pm 190$		& $1.84 \pm 0.23$ \\
	& \Lyalpha$_2$	& 0.544		& 12.1375		 & 12.1385		  & 0.0595		    & 1470			& 0.89 \\
\NeIX  	& He $r$	& 0.343		& 13.4473                & $13.4530\pm 0.0043$    & $0.0594 \pm 0.0098$	    & $1320 \pm 220$		& $1.54 \pm 0.32$ \\
	& He $i$	& 0.343		& 13.5531		 & 13.5589		  & 0.0599		    & 1320			& $0.38 \pm 0.19$ \\
	& He $f$	& 0.343		& 13.6990		 & 13.7048		  & 0.0605		    & 1320			& $1.00 \pm 0.27$ \\
\hline
\end{tabular}
\flushleft
$^\mathrm{a}$Results obtained just by fitting to MEG data \\
$^\mathrm{b}$Results obtained by omitting MEG~$-1$ data from the fit
\label{tab:LineFitting}
\end{table*}

\begin{figure*}
\centering
\includegraphics[width=5.5cm]{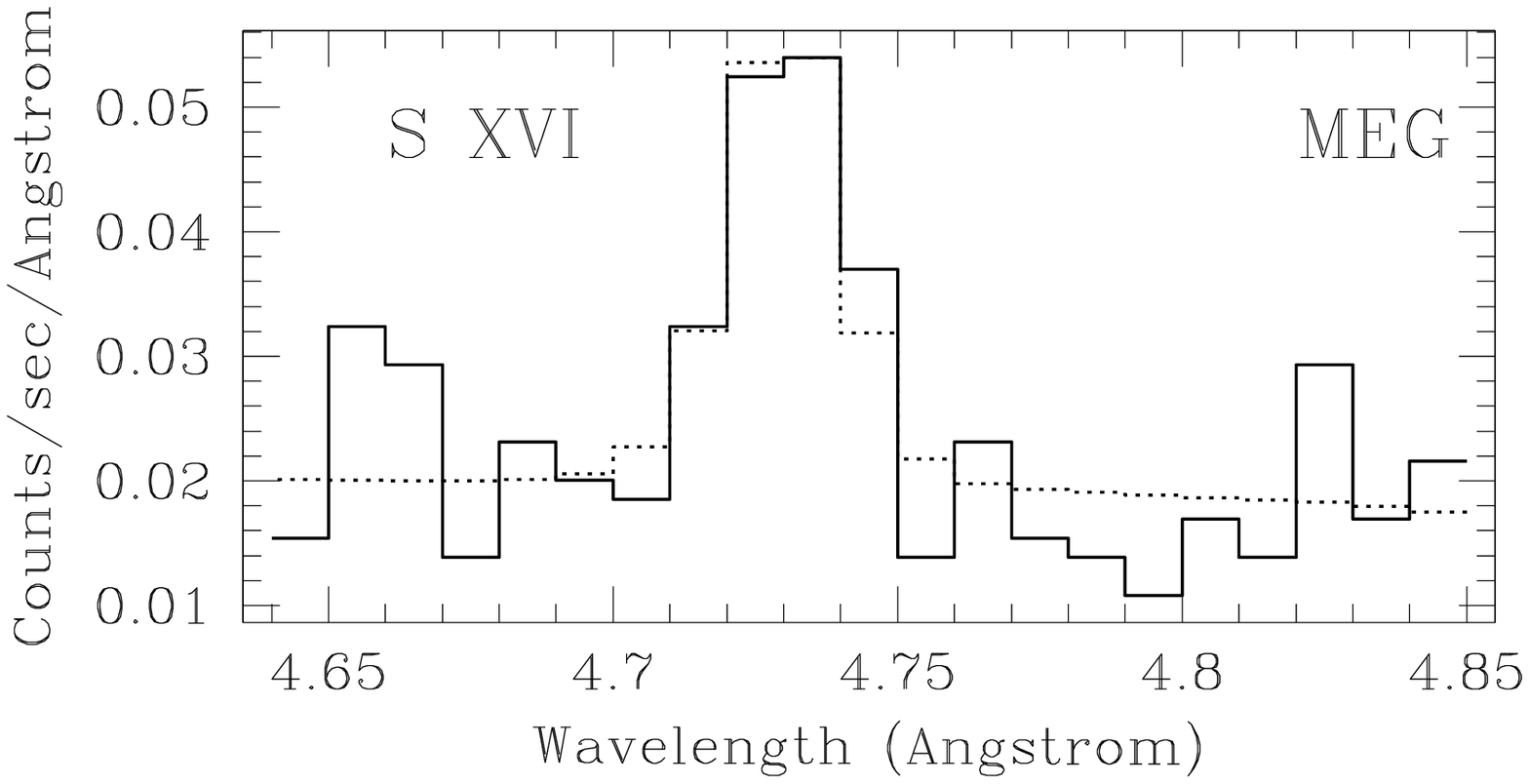}
\includegraphics[width=5.5cm]{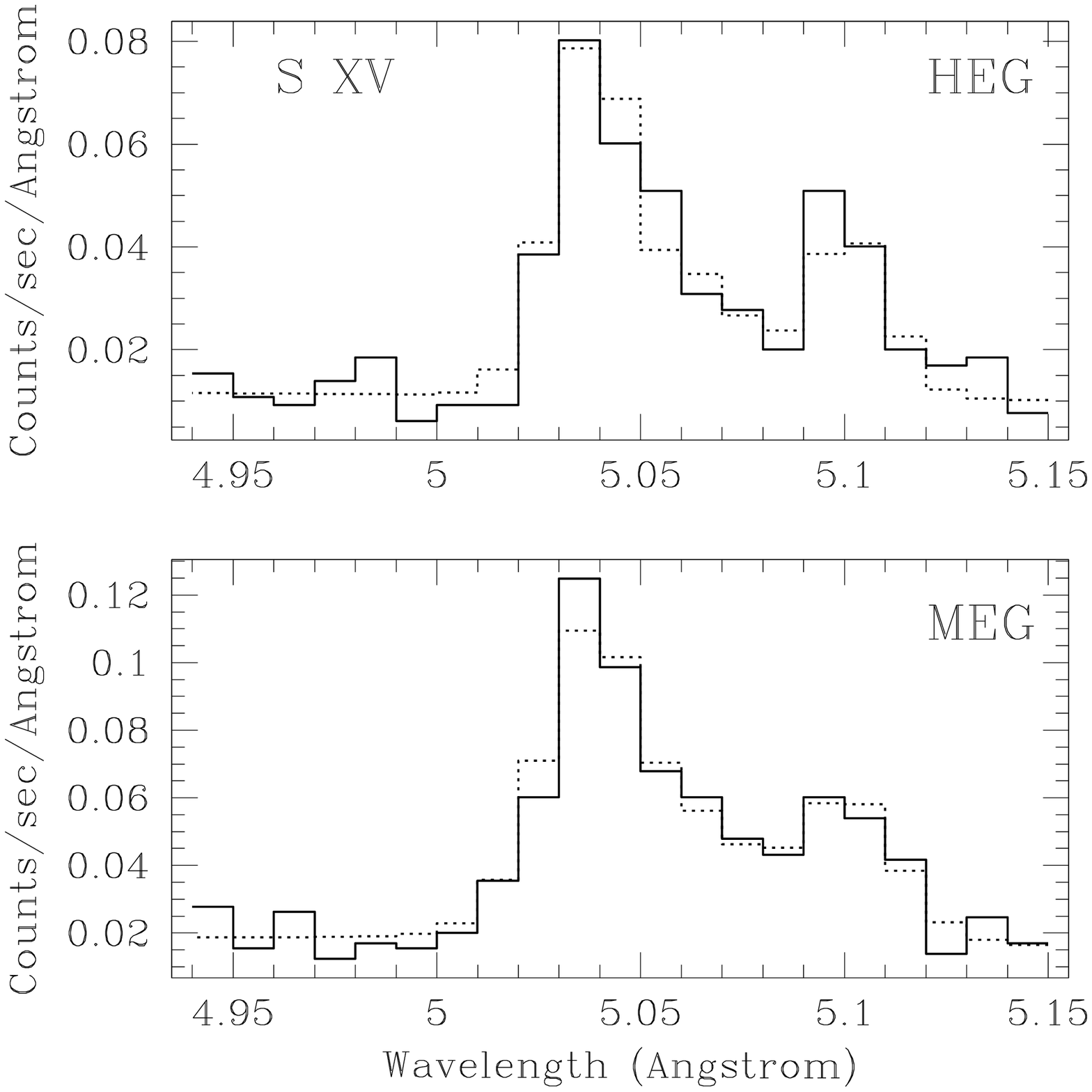}
\includegraphics[width=5.5cm]{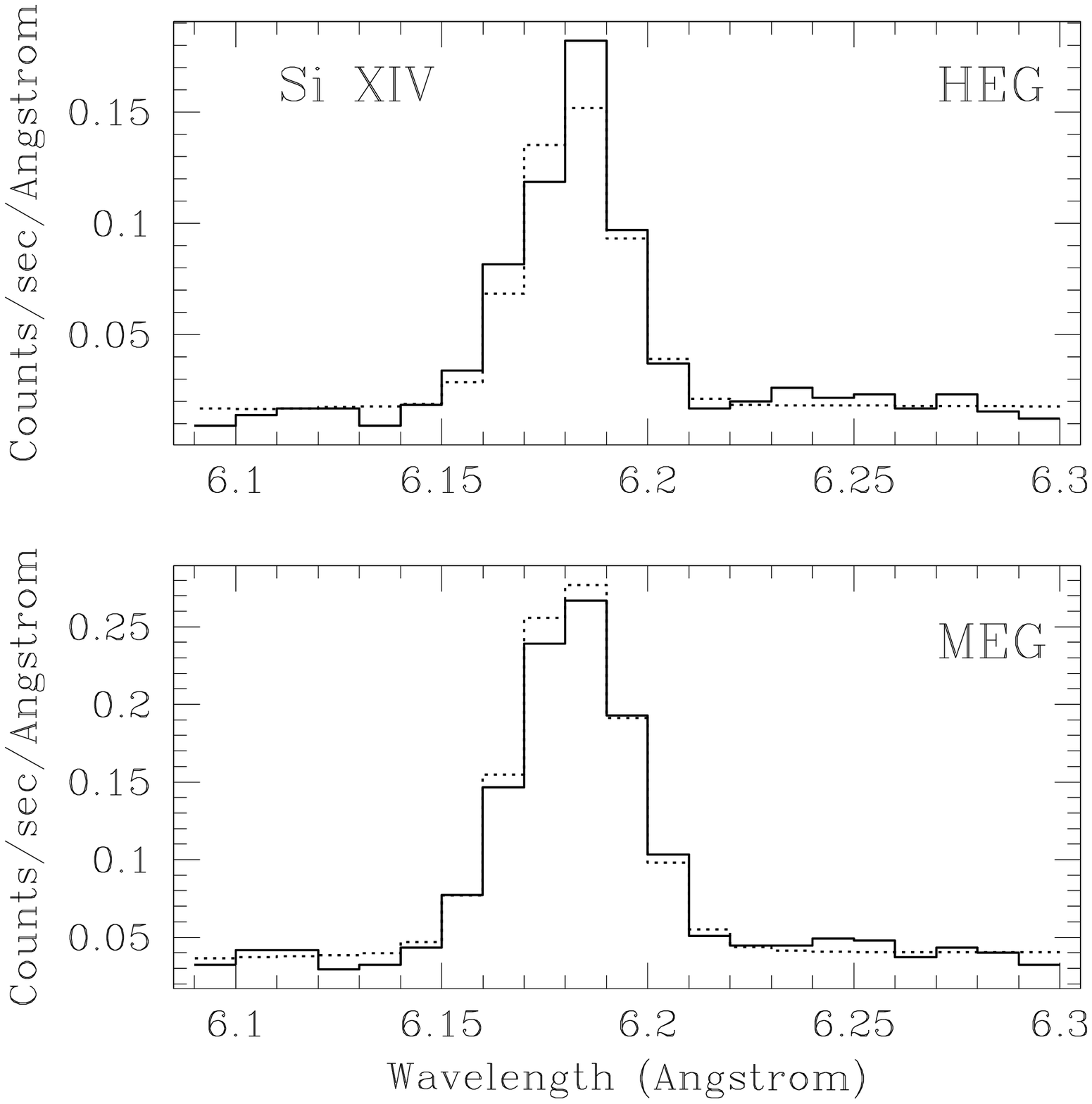} \\
\includegraphics[width=5.5cm]{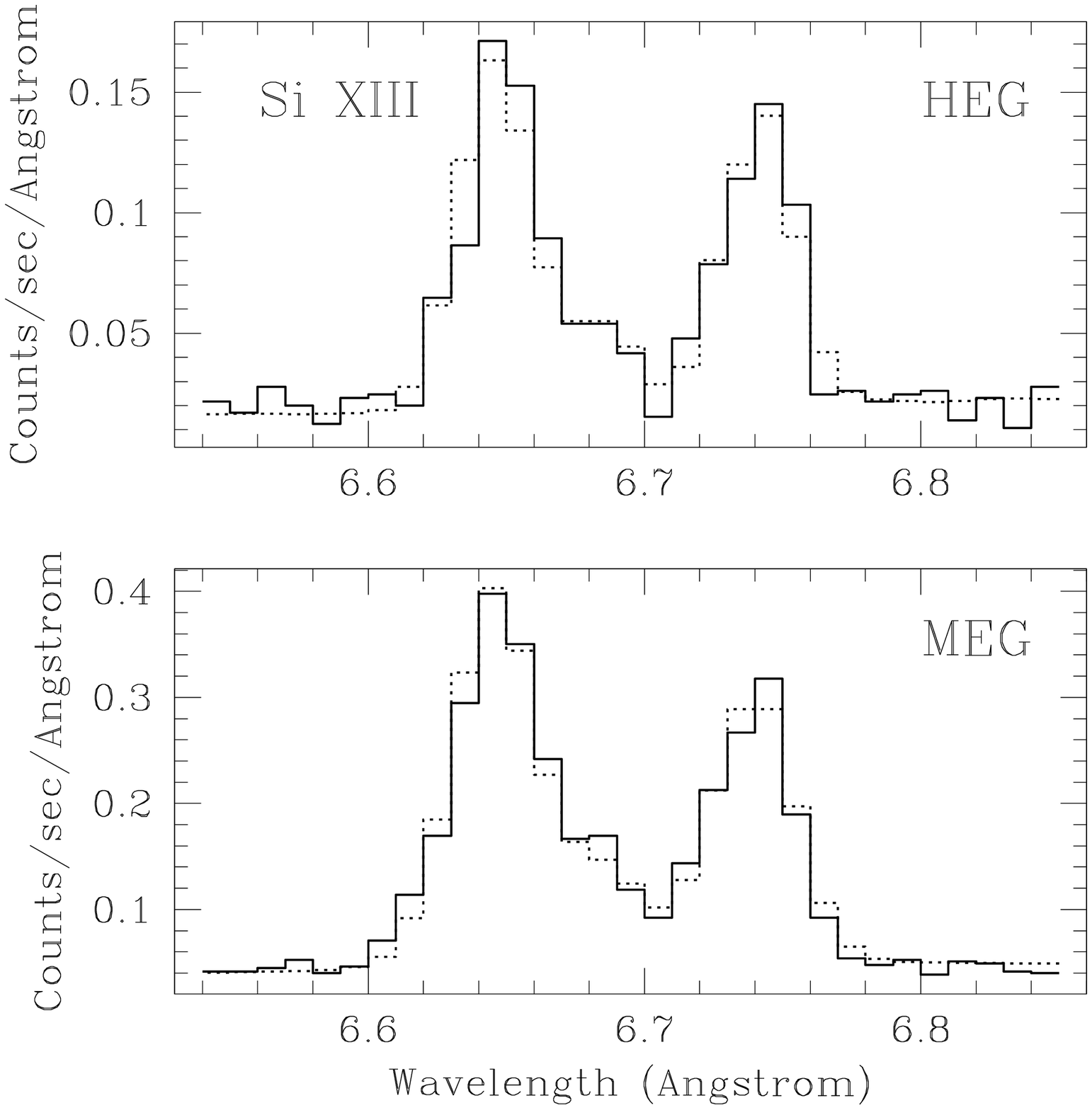}
\includegraphics[width=5.5cm]{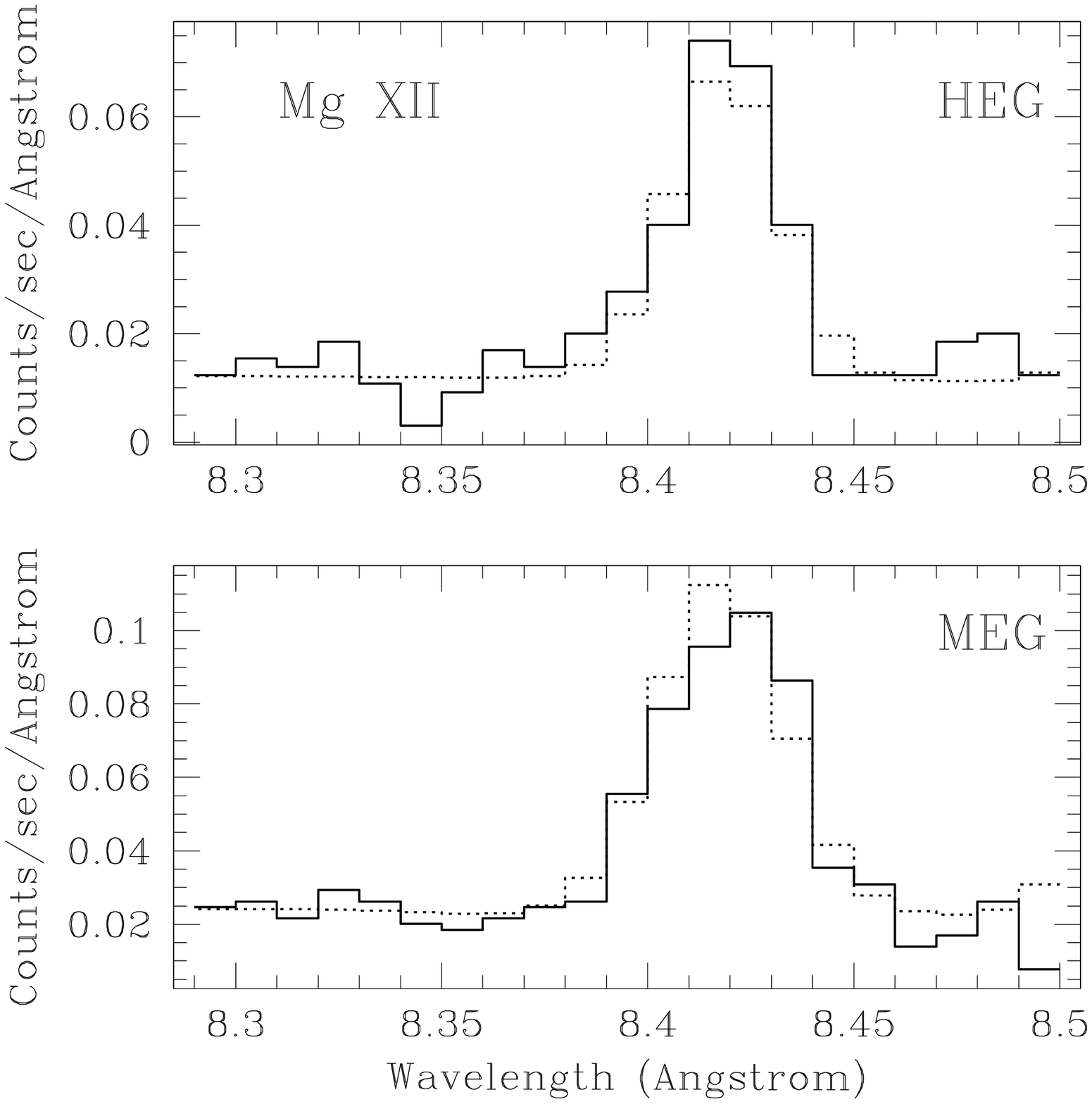}
\includegraphics[width=5.5cm]{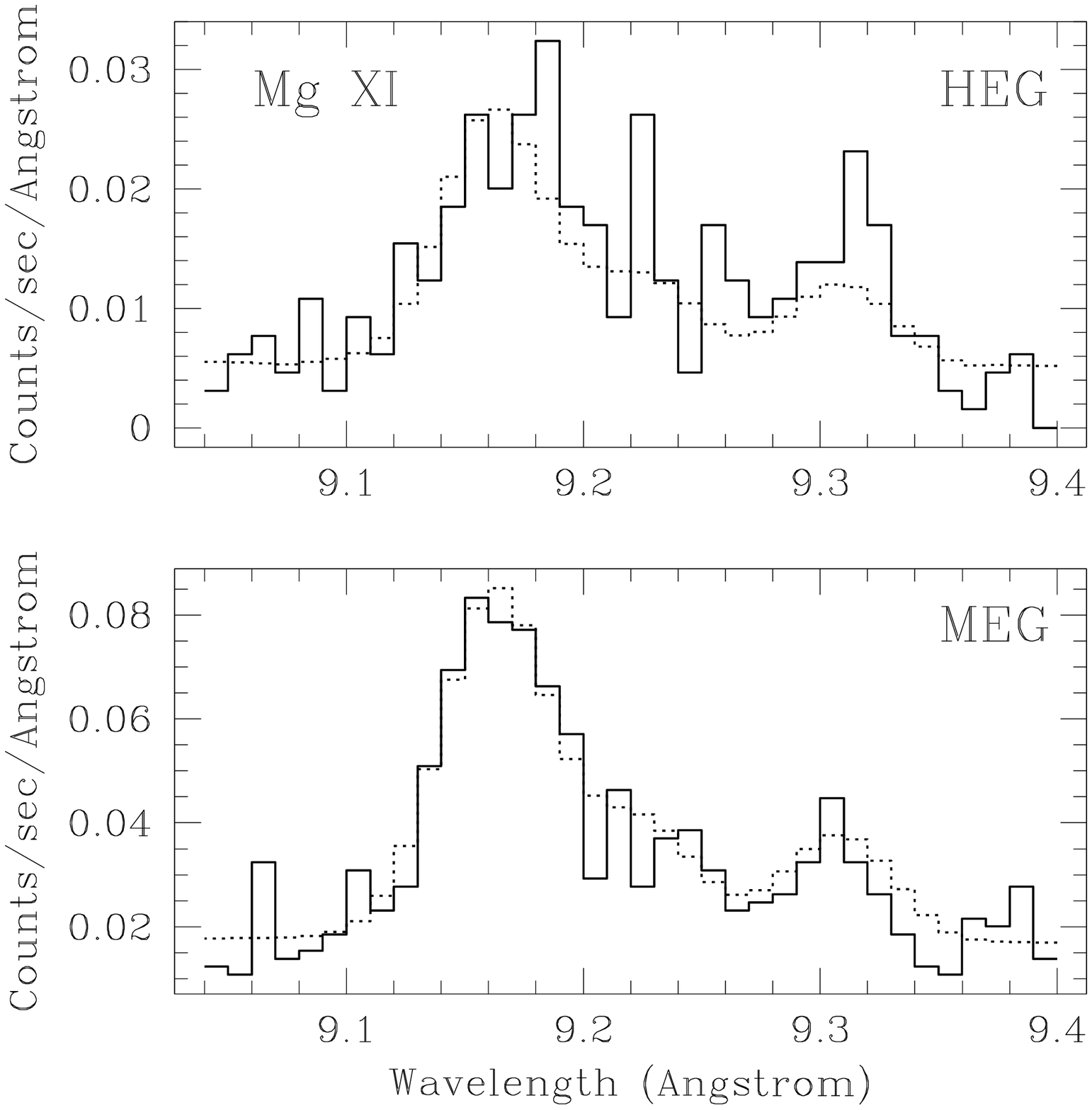} \\
\includegraphics[width=5.5cm]{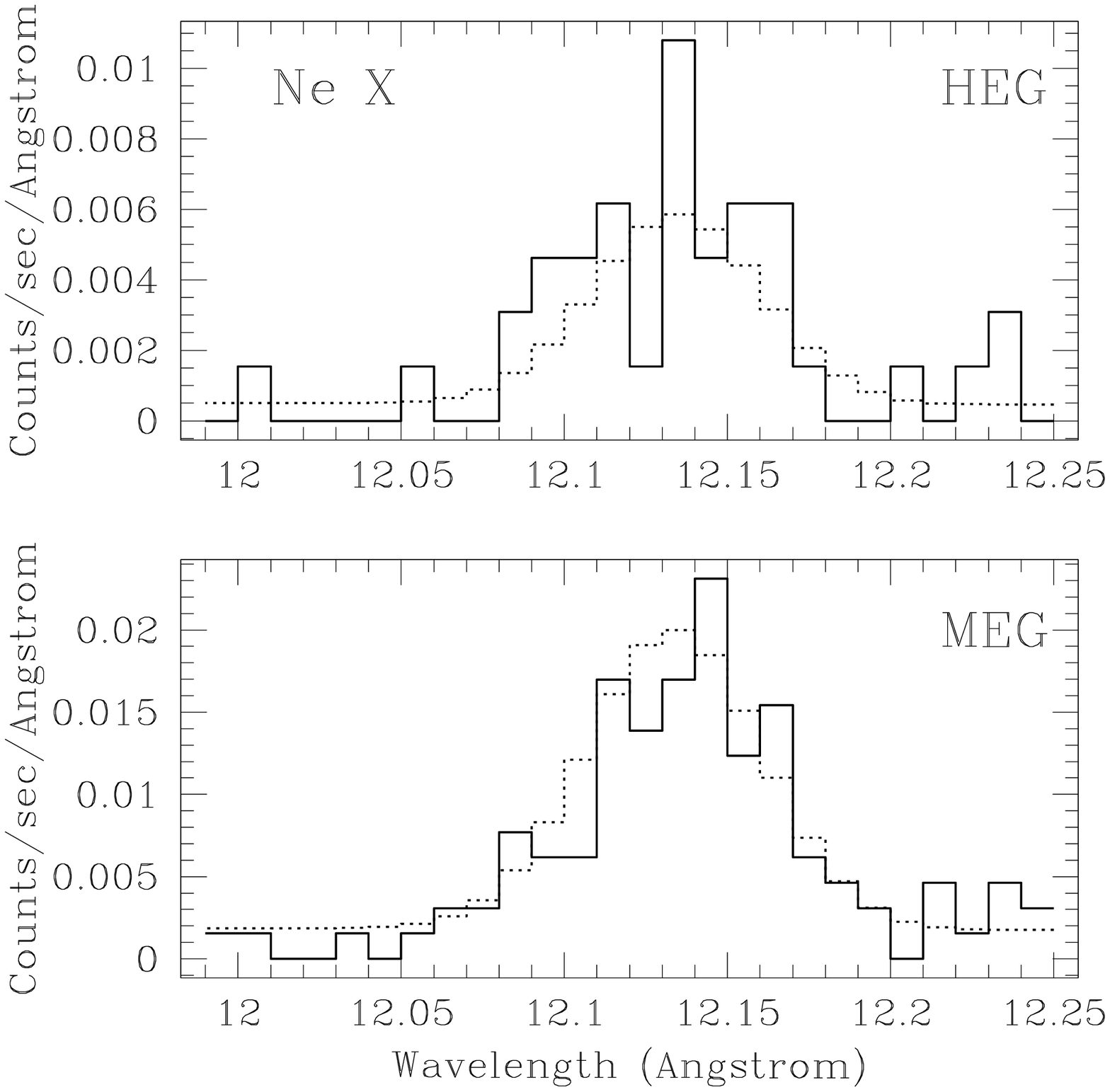}
\includegraphics[width=5.5cm]{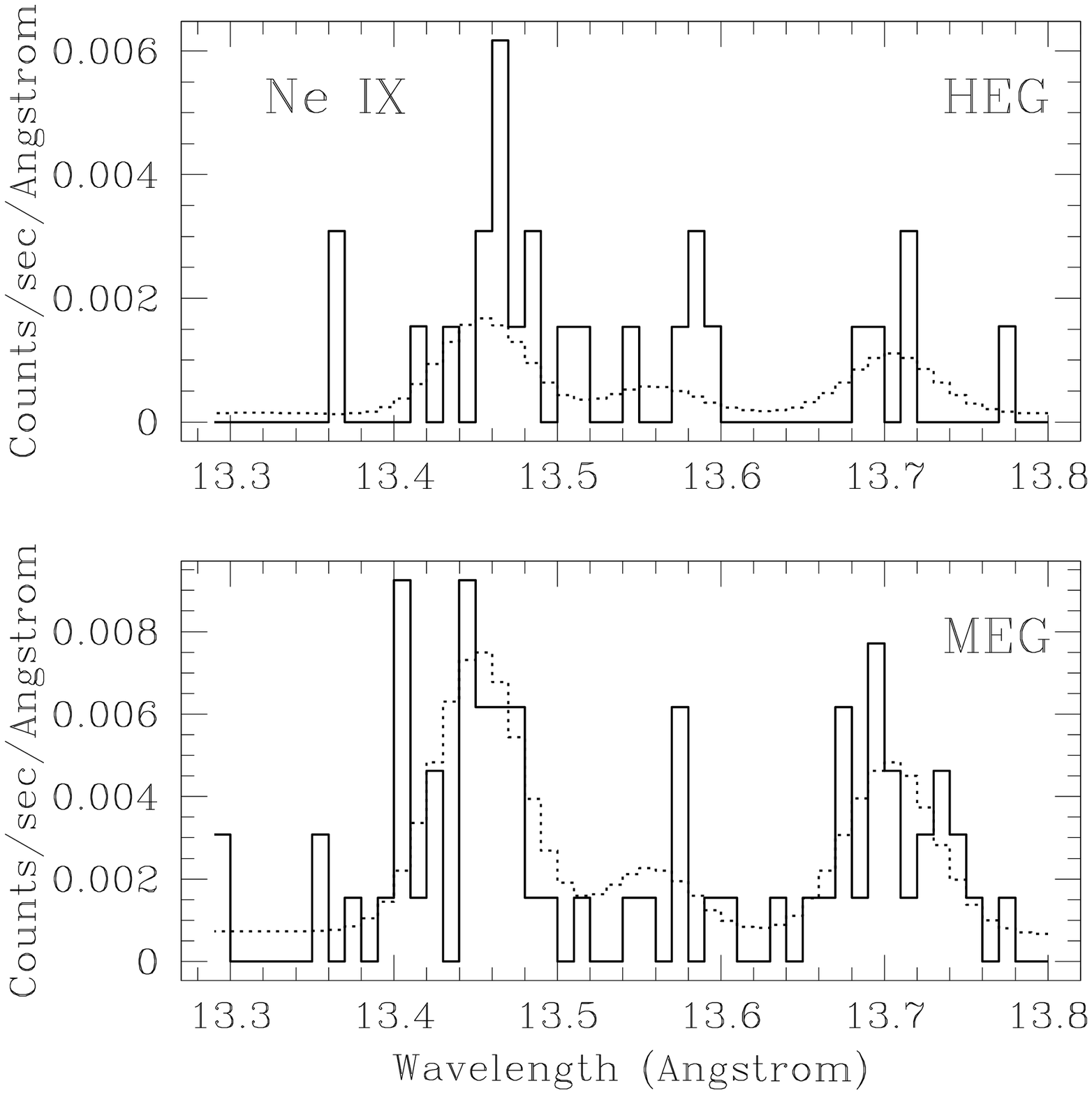}
\caption{The \Lyalpha\ lines from \SXVI, \SiXIV, \MgXII\ and \NeX\ and the He-like \fir\
triplets from \SXV, \SiXIII, \MgXI\ and \NeIX\ observed in the \chandra\ HETGS spectrum of $\gamma^2$~Vel.
The solid line in each panel shows the data, while the dashed
line shows the best fit model. For illustrative purposes only, the $+1$ and $-1$ orders of each grating
(HEG and MEG) have been co-added and binned up to 0.01\angstrom. Note that the \SXVI\ \Lyalpha\ fit was
just carried out on the MEG data, and so the HEG data are not plotted.}
\label{fig:LineMontage}
\end{figure*}

\begin{figure*}
\centering
\includegraphics[width=8cm]{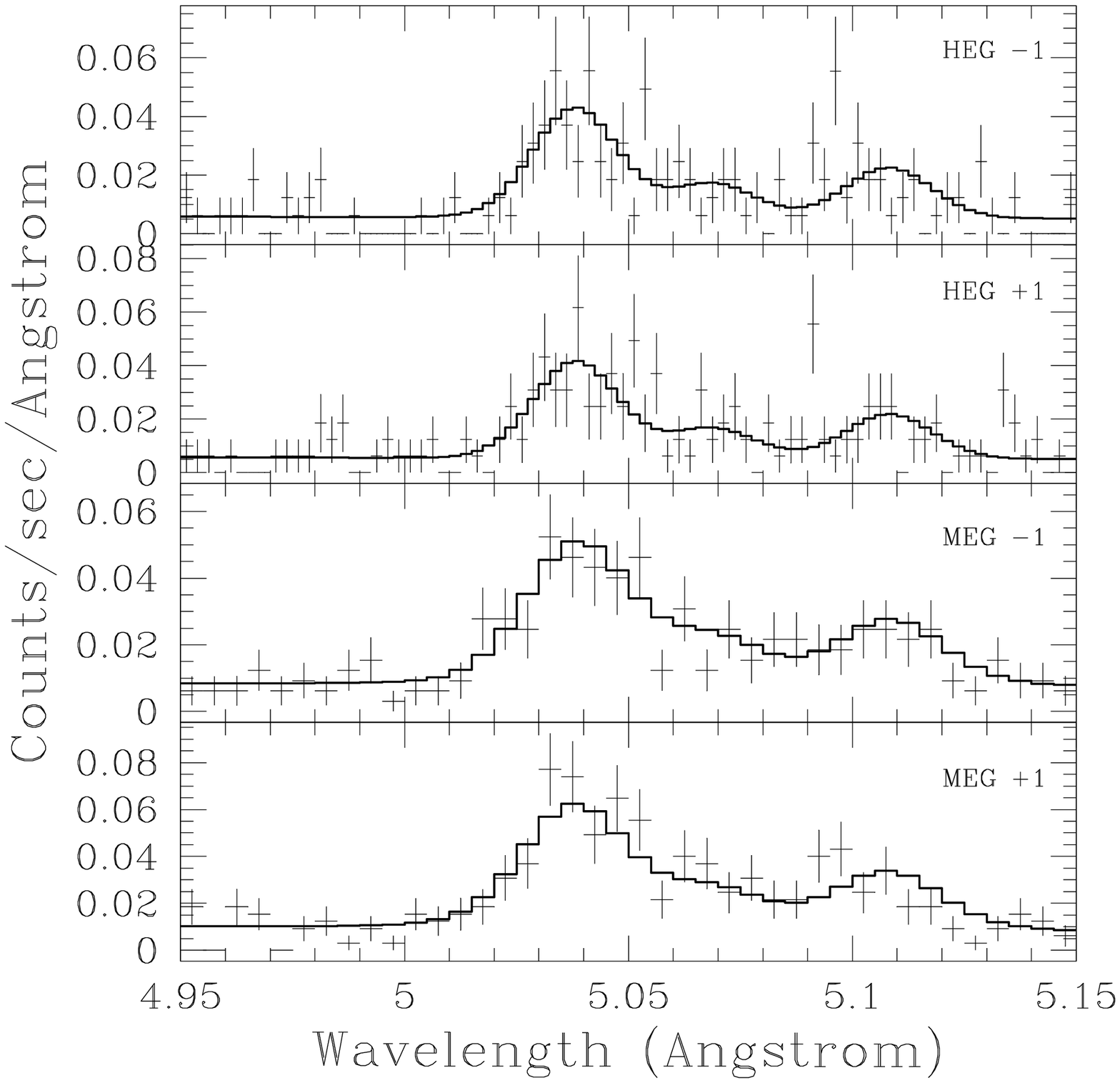}
\includegraphics[width=8cm]{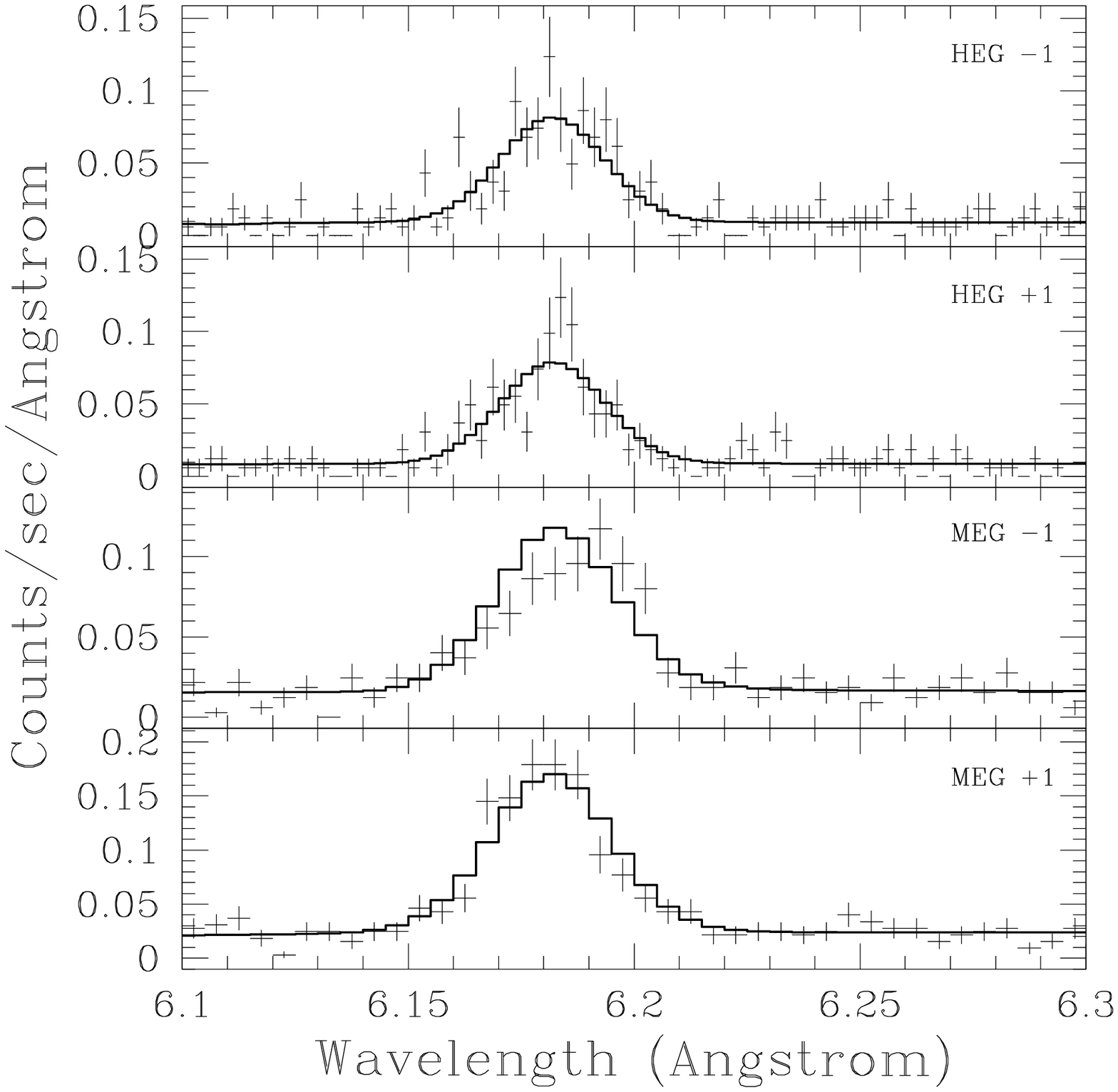}
\caption{The unbinned \SXV\ \fir\ triplet (left) and \SiXIV\ \Lyalpha\ line (right), with the
best fitting Gaussian line profiles. The anomalous appearance of the MEG~$-1$ line in the
right figure is discussed in the text.}
\label{fig:Unbinned}
\end{figure*}

The \Lyalpha\ lines and He-like \fir\ triplets from S, Si, Mg and Ne are shown in Fig.~\ref{fig:LineMontage},
along with the best-fitting model for each. For illustrative purposes only the data have been binned up to 0.01\angstrom\
in these plots. For comparison, the unbinned \SXV\ \fir\ triplet and \SiXIV\ \Lyalpha\ line are shown
in Fig.~\ref{fig:Unbinned}.
Note that there appears to be some excess emission in the \SXV\ \fir\ HEG data approximately 0.01\angstrom\ shortwards
of the forbidden line. However, this is probably not real, because it does not appear at the same wavelength in the
$+1$ and $-1$ orders, it does not appear at all in the MEG data, and there is no corresponding feature shortwards of
the resonance line. Note also that the MEG~$-1$ \SiXIV\ \Lyalpha\ line appears
significantly different from the others -- it is shifted towards the red, and also it appears to be skewed
redwards. However, this skewing is probably not real -- if it were it should be detected in the other three
spectra (if anything, the line in the MEG~$+1$ spectrum appears to be slightly skewed bluewards).
The wavelength obtained by fitting to just the MEG~$-1$ spectrum is $6.1834 \pm 0.0014 \angstrom$,
that obtained by fitting to the other three spectra is $6.1795 \pm 0.0006 \angstrom$, and that obtained
by fitting to all four spectra is $6.1803 \pm 0.0006 \angstrom$. Hence, while the MEG~$-1$ spectrum gives
a significantly different wavelength from the other three spectra, including it in the fit does not
significantly affect the results (this is true of the FWHM and flux as well). Nevertheless, as the line
is clearly different in the MEG~$-1$ spectrum, we have chosen to quote the results obtained by omitting
the MEG~$-1$ spectrum from the fit in Table~\ref{tab:LineFitting}

\begin{figure*}
\centering
\includegraphics[width=8cm]{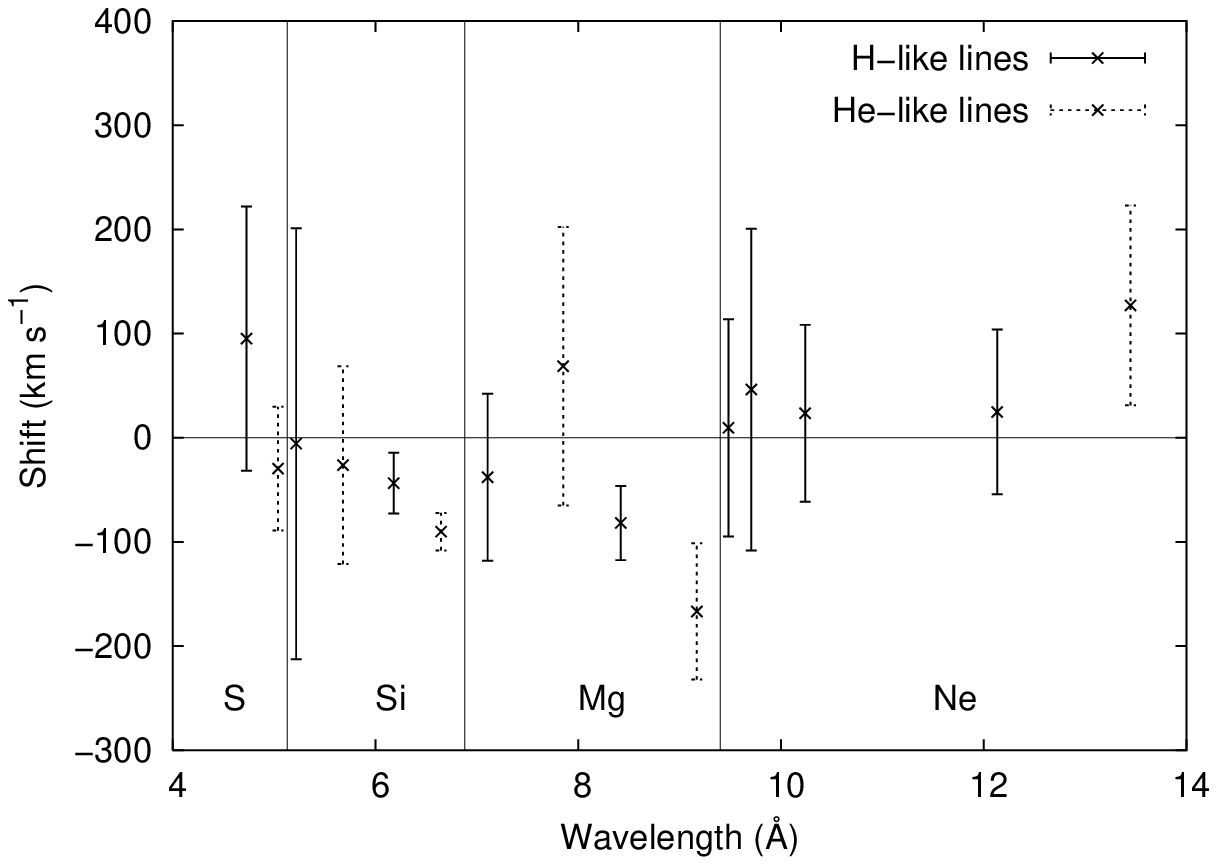}
\includegraphics[width=8cm]{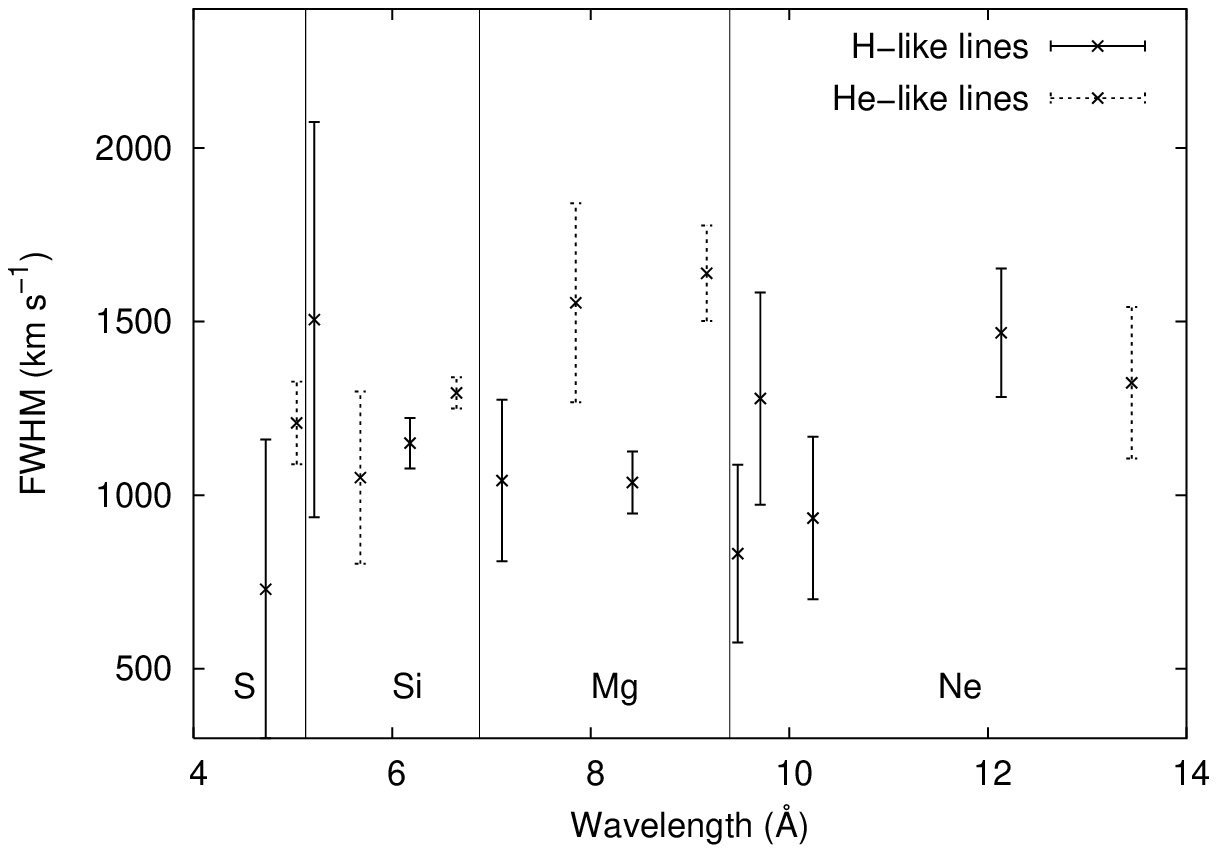}
\caption{Line shifts (left) and widths (right) as functions of laboratory wavelength. The vertical lines divide the plot
up into lines from S, Si, Mg and Ne (from left to right). Lines from H-like and He-like ions are denoted with
solid and dashed error bars, respectively.}
\label{fig:LineFittingResults}
\end{figure*}

The fit results are illustrated in Fig.~\ref{fig:LineFittingResults},
which shows the measured line shifts and widths as functions of laboratory wavelength.
One can see from the results that the lines are typically unshifted from their lab wavelengths,
with FWHMs of $\sim$1000--1500\kmps. The shifts and widths are uncorrelated with lab wavelength
or the ionization potential of the emitting ion.
The line shifts are well fit with a single mean shift (independent
of wavelength) of $-64 \pm 12 \kmps$ ($\rchisq = 1.13$ for 14 degrees of freedom), while the line widths
are adequately fit with a single mean FWHM of $1240 \pm 30 \kmps$ ($\rchisq = 1.91$ for 14 degrees of freedom).

\begin{figure}
\centering
\includegraphics[width=8cm]{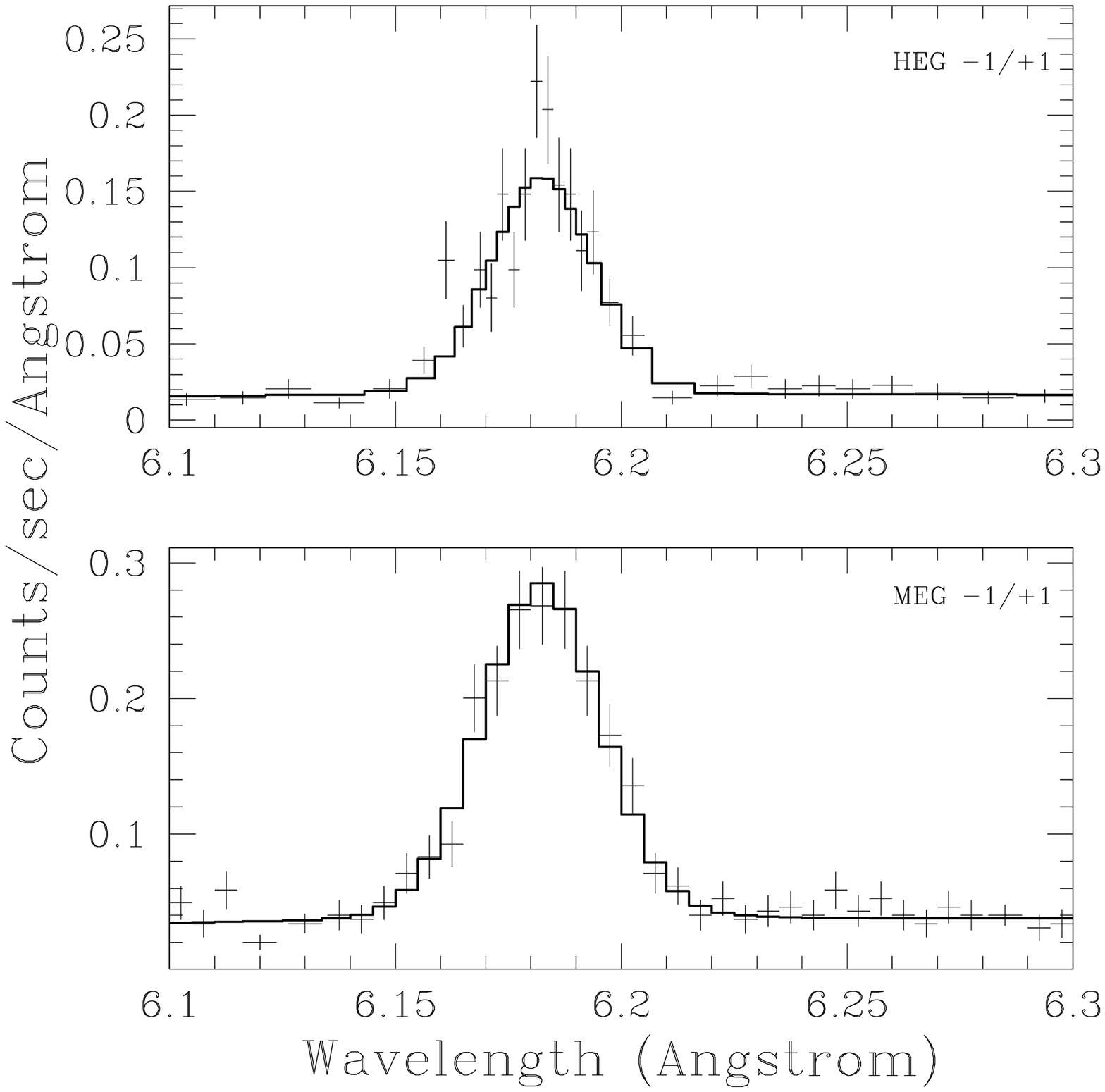}
\caption{The \SiXIV\ \Lyalpha\ line shown binned up so there are at least 10 counts per bin, with
the best fit Gaussian line profile, obtained by fitting to the two binned spectra simultaneously
using $\chi^2$.}
\label{fig:Si14LyaBinned}
\end{figure}

While the Cash statistic can be used on data with low numbers of counts, it does not contain any goodness-of-fit
information. Therefore, to get some idea of the goodness of the fits, we again co-added the $+1$ and $-1$ orders
of the HEG and MEG spectra, binned up the data so there were at least 10 counts per bin, and then repeated
the fitting procedure using $\chi^2$ (which does enable us to assess the goodness of fit). Just for this $\chi^2$
fitting, the HEG~$-1$ and MEG~$+1$ RMFs were used, as in the broad-band spectral fitting.
To illustrate the effect of rebinning the data, the binned \SiXIV\ \Lyalpha\ line is shown in
Fig.~\ref{fig:Si14LyaBinned}. 
We found no significant difference in the wavelengths, widths and fluxes derived using the two different fit statistics --
all agreed within $1 \sigma$. In all cases, Gaussians give good fits to the emission lines. A possible exception
to this is the \SiXIII\ \fir\ triplet -- Gaussians do not give good fits to the HEG lines (which appear to be
slightly skewed redwards) but they do to the MEG lines. However, this skewing is probably not real, as there is
no evidence for it in the MEG data (which has higher signal-to-noise). There is therefore no convincing evidence
that Gaussian line profiles do not give good fits to the observed emission lines.

There is no improvement in $\chi^2$ when two Gaussians are used for the Lyman lines as opposed to just one.
However, while it is not required by the data, the use of a doublet model is justified on physical grounds,
as discussed above.

\subsection{Comparison with broad-band results}
\label{subsec:Comparison}

\begin{figure*}
\centering
\includegraphics[width=8cm]{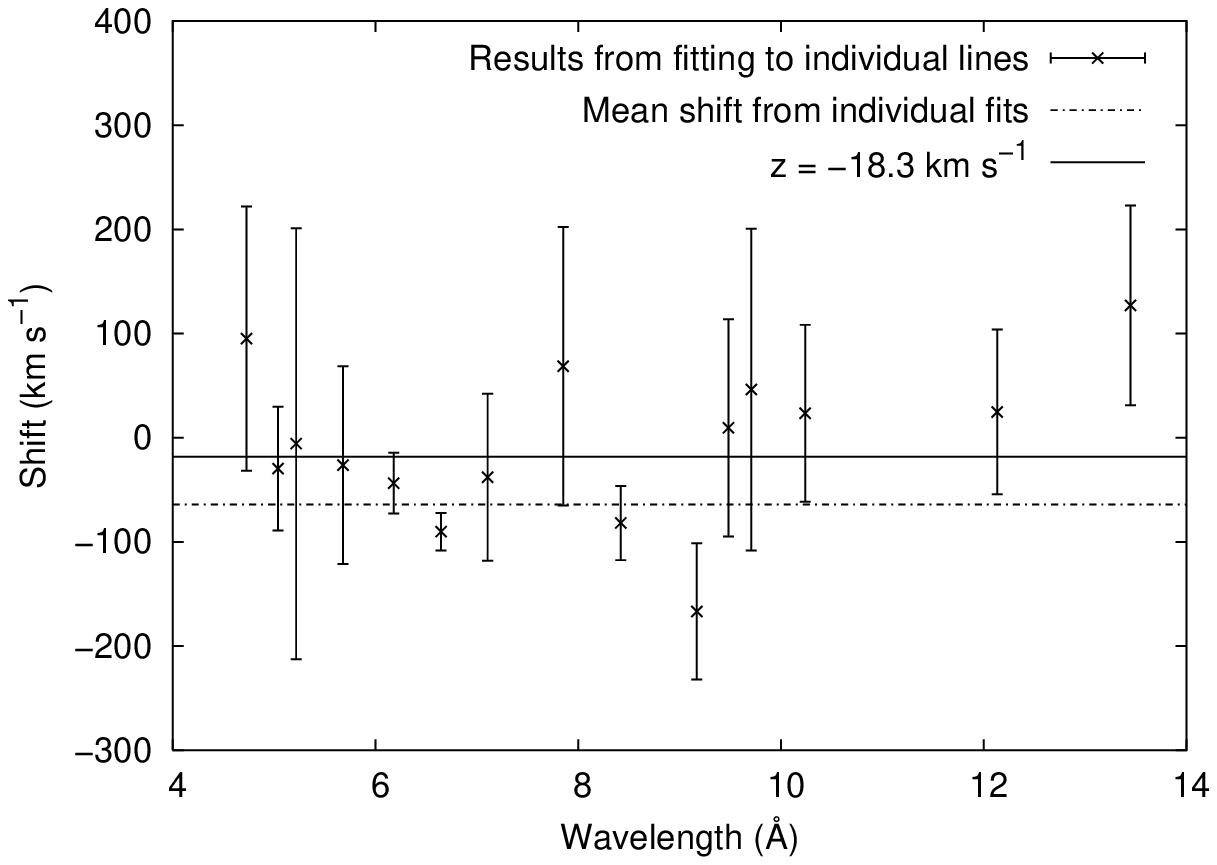}
\includegraphics[width=8cm]{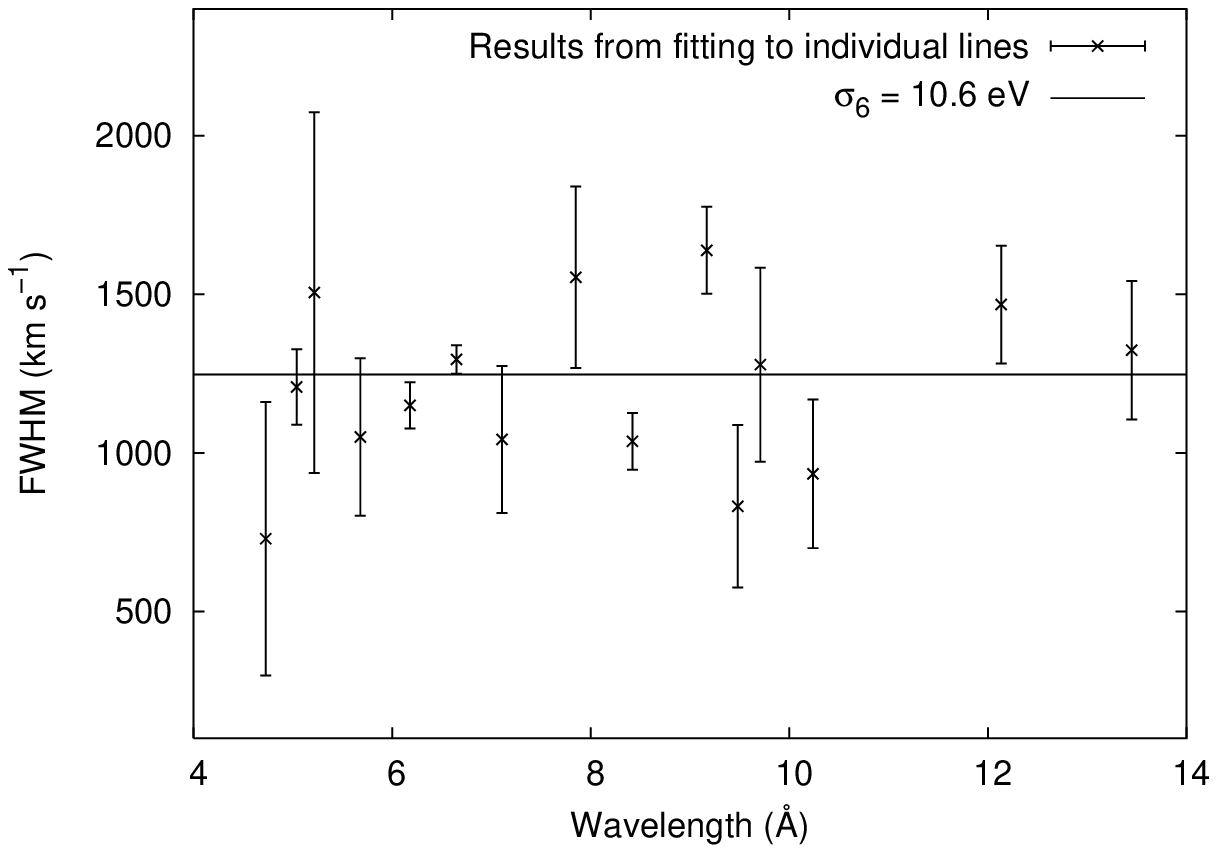}
\caption{Comparison between line shifts (left) and line widths (right) derived from fitting
to the broad-band spectrum and those derived from fitting to individual emission lines.
The data points in both figures are the results from fitting to individual emission lines.
The dot-dashed line in the top figure shows the mean shift of the lines ($-64\kmps$).
The solid line in the top figure shows the line shift corresponding to $z = -18.3\kmps$,
and the solid line in the bottom figure shows the line width corresponding to
$\sigma_6 = 10.6\ev$ (model B, Table~\ref{tab:Broadband}).}
\label{fig:BroadbandComparison}
\end{figure*}

The results described in this section can be compared with the results derived from the
broad-band spectrum in Section~\ref{subsec:2Tapec}. Fig.~\ref{fig:BroadbandComparison} compares
the line shifts and widths derived in this section with those inferred from the values of
$z$ and $\sigma_6$ in Table~\ref{tab:Broadband} (model B). There is excellent agreement
between the width derived from the broadband fitting (FWHM = $1250^{+20}_{-50} \kmps$)
and the mean width of the individual emission lines (FWHM = $1240 \pm 40 \kmps$). Furthermore,
the lack of any observed correlation between line width and wavelength justifies our use
of $\alpha = 1$ in equation~(\ref{eq:gsmooth}).

We would expect there to be good agreement between the line shifts derived from the broadband fitting 
and the mean shift of the individual emission lines, as the rest wavelengths in the \texttt{apec} model
used to measure the former are the same as those used to calculate the latter. In fact, the line shift
derived from the broadband fitting ($z= -18.3^{+0.9}_{-1.5} \kmps$) disagrees with the mean shift of the
individual emission lines ($-64 \pm 12 \kmps$) at the $3 \sigma$ level. However, the agreement is probably
better than this indicates, because the poor spectral fits in Section~\ref{sec:Broadband} mean the errors on
the spectral parameters (including $z$) are likely to be underestimated.

\subsection{Estimating the temperature using line flux ratios}
\label{subsec:TempLineRatios}

There are two simple ways to estimate the temperature $T$ of the X-ray-emitting plasma using the
flux ratios of emission lines. The first uses the flux ratio of the H-like \Lyalpha\
line to the He-like resonance line for a given element. As the temperature increases the
ionization balance shifts from He- to H-like ions, and this change is
manifested in the line flux ratio. The second uses the flux ratio $G = (f + i) /r$ of the
resonance ($r$; 1s2p~$^1$P$_1$ $\rightarrow$ 1s$^2$~$^1$S$_0$),
intercombination ($i$; 1s2p~$^3$P$_{1,2}$ $\rightarrow$ 1s$^2$~$^1$S$_0$) and
forbidden ($f$; 1s2s~$^3$S$_1$ $\rightarrow$ 1s$^2$~$^1$S$_0$) lines of a He-like ion. $G$
is a decreasing function of temperature \citep[e.g.][]{porquet01,paerels03},
due to the different temperature dependencies of the mechanisms for populating the upper levels of the lines
(direct collisional excitation from the ground state for the resonance line, cascades from higher levels populated
by dielectronic recombination for the forbidden and intercombination lines; see \citealp{smith01}).
The $G$ ratio is also sensitive to electron density at high densities, due to population of the upper level of the resonance
line from the 1s2s~$^1$S$_0$ level \citep{porquet01}. However, the densities at which this becomes important
\citep[$\Ne \ga 10^{12} \pcc$;][]{porquet01} are larger than those expected in the wind-wind collision in $\gamma^2$~Vel
(the hydrodynamical simulations in Section~\ref{sec:Modelling} indicate $\Ne \la 10^{11} \pcc$).

To estimate temperatures using H-like to He-like flux ratios, we use line emissivities calculated as functions of $T$ with
the Astrophysical Plasma Emission Code and Database
\citep[\textsc{apec}/\textsc{aped}\footnote{http://cxc.harvard.edu/atomdb/};][]{smith00,smith01}.
These emissivities are distributed in \textsc{atomdb} v1.1.0.
Comparing the measured H-like to He-like flux ratios to the calculated values, we infer temperatures of $5.6 \pm 0.5 \MK$
for Ne, $8.6 \pm 0.3 \MK$ for Mg, $11.7 \pm 0.2 \MK$ for Si and $14 \pm 1 \MK$ for S.

The measured $G$ ratios are $0.9 \pm 0.3$, $0.67 \pm 0.10$ and $0.91 \pm 0.05$ for \NeIX, \MgXI\ and \SiXIII,
respectively. To calculate temperatures, we compare these values to the values calculated by \citet{porquet01}
in the low-density limit. Their calculations include the effects of blending of dielectronic satellite lines.
The contribution of these blended lines to the measured fluxes depends on the spectral resolution of
the instrument used \citep{porquet01}, and so they tabulate $G$ ratios as a function of $T$ assuming the
MEG spectral resolution [$\Delta \lambda (\mathrm{FWHM}) = 23 \mA$] and assuming the HEG spectral resolution
($\Delta \lambda = 12 \mA$). Our fits use data from the MEG and the HEG simultaneously. However, this does
not matter as there are no significant differences between the temperatures inferred using the MEG values
or the HEG values from \citet{porquet01}. Using the MEG values from \citet{porquet01} we infer temperatures of
$3^{+3}_{-2} \MK$ for \NeIX, $7^{+3}_{-2} \MK$ for \MgXI\ and $4.8^{+0.8}_{-0.7} \MK$ for \SiXIII, whereas using
the HEG values we infer $2^{+4}_{-2} \MK$ for \NeIX, $7^{+3}_{-2} \MK$ for \MgXI\ and $5.4^{+0.8}_{-0.7} \MK$ for \SiXIII.

The temperatures measured using the \fir\ triplets of \NeIX\ and \MgXI\ are in good agreement with those measured
using the H-like to He-like flux ratios. However, the temperature measured from the \SiXIII\ \fir\ triplet is
significantly cooler than that measured from the \SiXIV:\SiXIII\ flux ratio. The calculation of the temperature from the $G$
ratio may be affected by blending of the \SiXIII\ forbidden line with \MgXII\ \Lygamma. This would mean the $G$ ratio is
over-estimated and $T$ is under-estimated by this method, in agreement with the observed discrepancy. However,
the discrepancy may more simply be due to the He-like emission coming from different, cooler regions than the H-like emission.

\subsection{The location of the X-ray emitting plasma}
\label{subsec:Location}

The flux ratio $R = f/i$ is sensitive both to electron density \Ne\ and the UV radiation field, due to depopulation of the
upper level of the forbidden line to the upper level of the intercombination line \citep{gabriel69,porquet01,paerels03}.
Comparing the measured $R$ ratios with theoretical values gives information on the electron density
and/or UV radiation field, and with some other assumptions the location of the X-ray emitting plasma can be inferred.

The measured $R$ ratios are $2.6 \pm 1.5$ for Ne, $1.0 \pm 0.2$ for Mg and $2.5 \pm 0.2$ for Si. \citet{porquet01}
have calculated $R$ ratios as functions of \Ne\ for a range of electron temperatures, radiation temperatures and
radiation dilution factors [$W(r) = 0.5(1-\sqrt{1-(\Rstar/r)^2})$, where $r$ is the distance from the centre of a star
of radius \Rstar]. The radius of the WR star is 3.2\Rsol\ \citep{demarco00}, and the distance from the WR star to the
wind-wind interaction will be at least half the separation (i.e. $r > 100\Rsol$). The dilution factor of the WR star's
radiation field is thus $W < 0.00026$, and its effect on the observed $R$ ratios will be negligible.

\begin{figure}
\centering
\includegraphics[width=8cm]{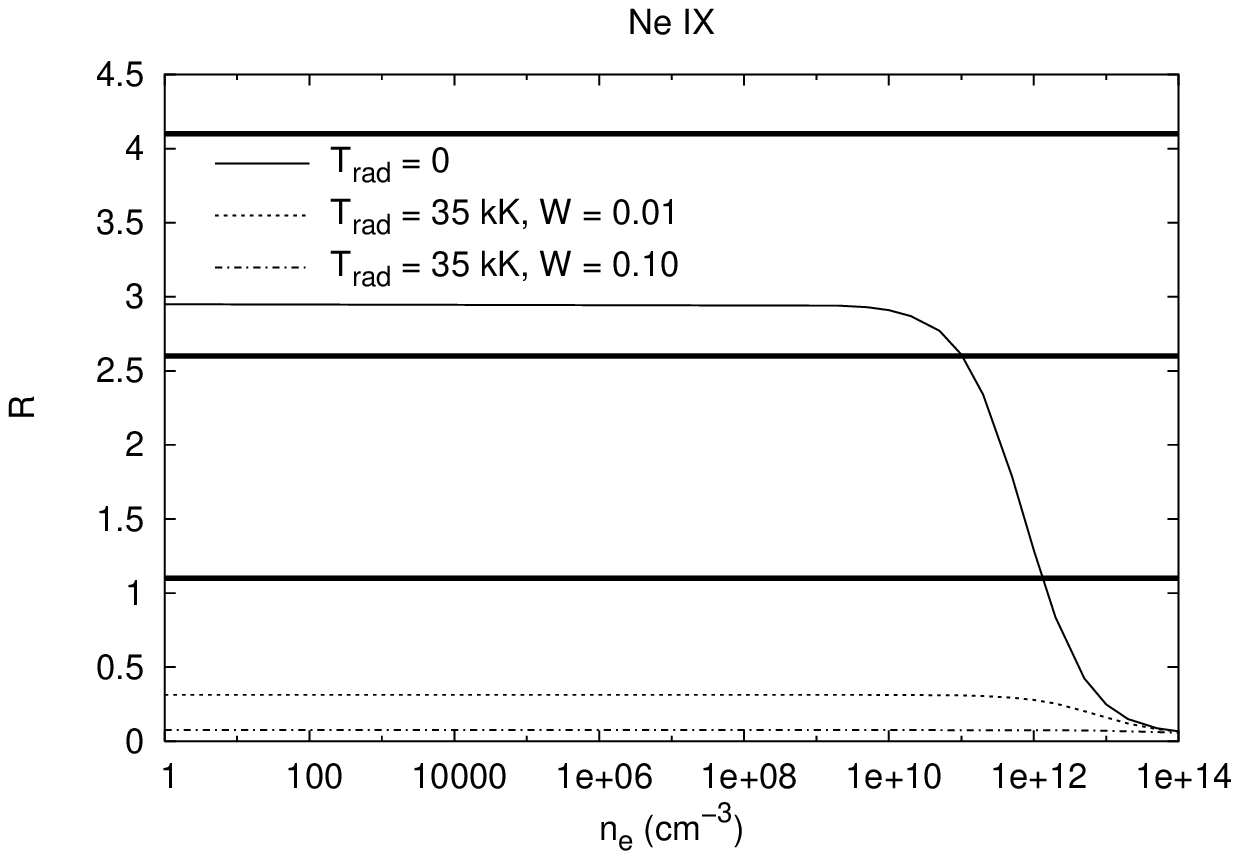}
\includegraphics[width=8cm]{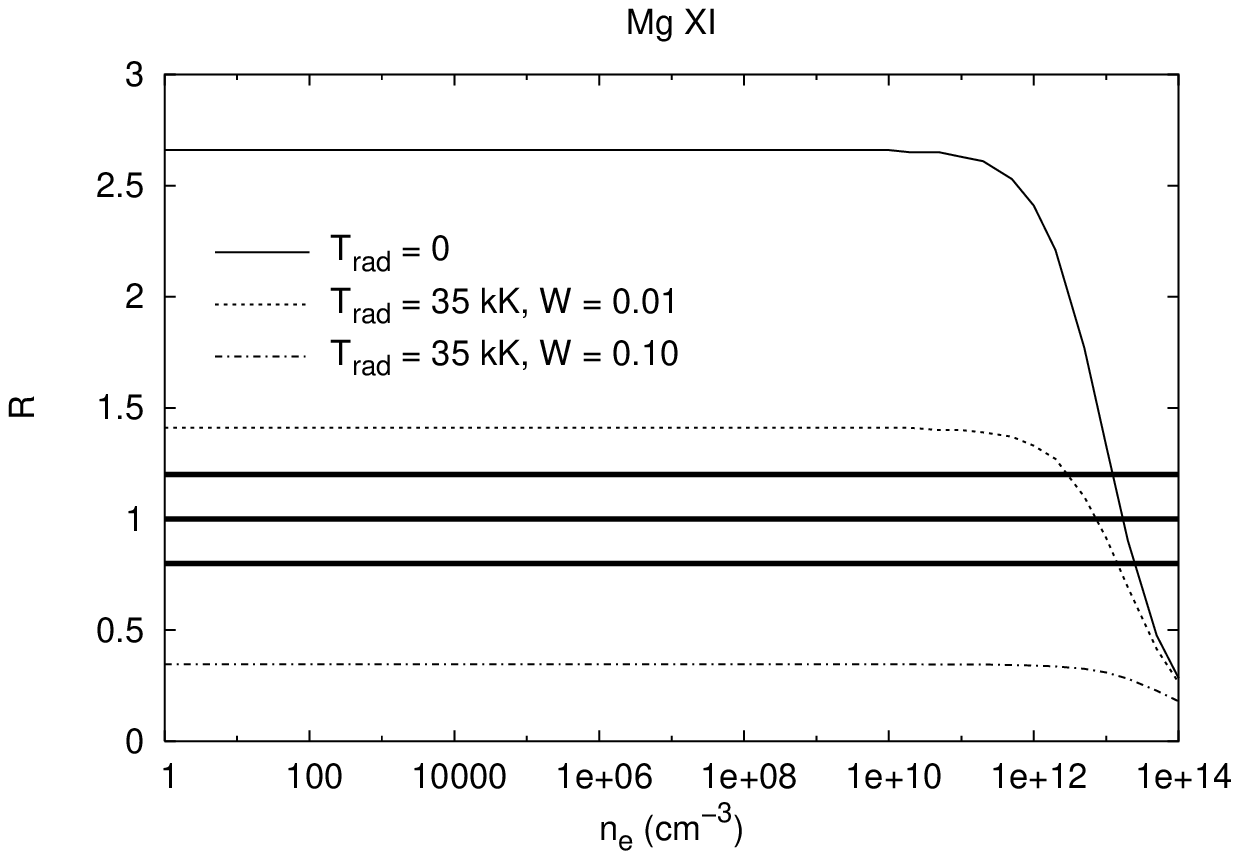}
\includegraphics[width=8cm]{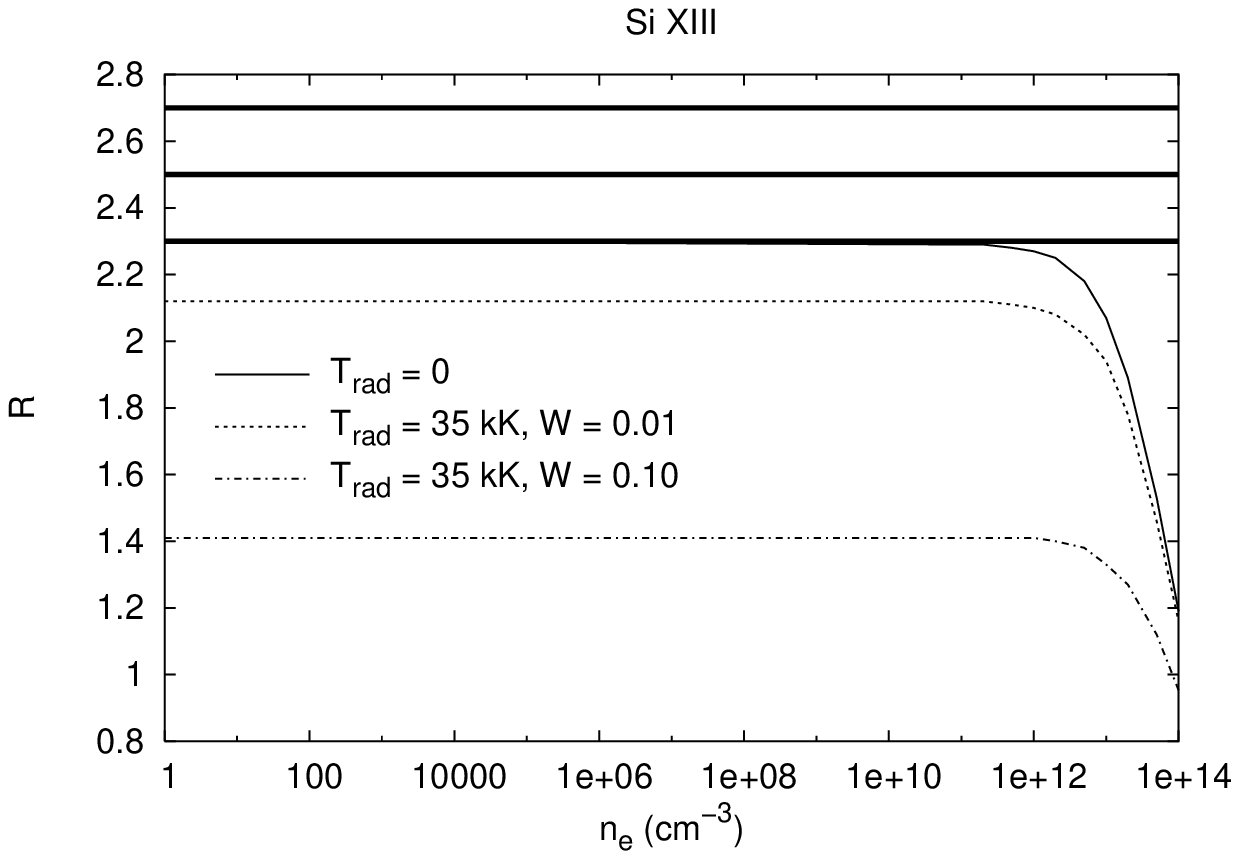}
\caption{Measured and calculated $R$ ratios for \NeIX, \MgXI\ and \SiXIII. The thin curves show the $R$ ratio as a function of
\Ne\ for different radiation temperatures ($T_\mathrm{rad}$) and dilution factors ($W$).
The thick horizontal lines show the measured $R$ ratios and the $1\sigma$ errors.}
\label{fig:Rratio}
\end{figure}

Fig.~\ref{fig:Rratio} shows the $R$ ratio as a function of \Ne\ in the absence of a radiation field ($T_\mathrm{rad} = 0$),
and in the presence of a blackbody radiation field with $T_\mathrm{rad} = 35\,000\K$
\citep[the effective temperature of the O star;][]{demarco99} for two different values of the dilution factor,
calculated by averaging the data for 30\,000\K\ and 40\,000\K\ in \citet{porquet01}. 
The plotted curves are for electron temperatures of 3.0\MK\ (\NeIX), 6.3\MK\ (\MgXI)
and 5.0\MK\ (\SiXIII). These temperatures were chosen from the values tabulated by \citet{porquet01} based on the
values derived in using $G$ ratios in Section~\ref{subsec:TempLineRatios}, though in fact the choice of electron temperature
has no effect on the conclusions. Also shown in Fig.~\ref{fig:Rratio} are the measured $R$ ratios.

For \NeIX\ the observed value of the $R$ ratio is consistent with the theoretical low-density, low-UV flux value
($R \approx 3$). Furthermore, we can see that even a relatively weak UV flux ($W \sim 0.01$) would
give rise to an $R$ ratio significantly lower than that which is observed. We therefore conclude
that $W \ll 0.01$. The radius of the O star is 12.4\Rsol\ \citep{demarco99}, and hence for the
\NeIX-emitting plasma, $r \gg 5\Rstar(\mathrm{O}) = 4 \times 10^{12} \cm$.
By the same argument, we can see from Fig.~\ref{fig:Rratio} that $W < 0.01$ for \SiXIII, and so we have
$r > 5\Rstar(\mathrm{O})$ for the \SiXIII-emitting plasma.

For \MgXI\ the observed value of the $R$ ratio is significantly lower than the theoretical low-density, low-UV flux
value ($R \approx 2.7$). If the UV radiation field were negligible, this would imply an electron density
of $\sim$$10^{13} \pcc$, which is far larger than is expected in the wind-wind collision zone. Even
with a dilution factor of $W \sim 0.01$, the implied density ($\sim$$10^{12}$--$10^{13} \pcc$) is too large.
On the other hand, if the dilution factor were as large as $W \sim  0.10$, the value of the $R$ ratio would be significantly
lower than that which is observed. Hence for the \MgXI-emitting plasma, $0.01 < W < 0.10$, and
so $1.7\Rstar(\mathrm{O}) < r < 5\Rstar(\mathrm{O})$. This means the \MgXI\ emission emerges from closer to the O star
than the \SiXIII\ emission. The implications of this are discussed in Section~\ref{sec:Discussion}.
It should be noted, however, that the uncertainty in the O star's emergent photospheric flux at the far- and
extreme-ultraviolet wavelengths relevant for the photoexcitation transitions (1270\angstrom\ for \NeIX,
1033\angstrom\ for \MgXI\ and 864\angstrom\ for \SiXIII; \citealp{porquet01}) may be significant, and that a
35\,000\K\ blackbody does not represent the flux of the O star very well, especially at the wavelengths
relevant to the higher $Z$ elements. A more accurate model of the photospheric flux will likely modify the
curves in Fig.~\ref{fig:Rratio} in a way that is not easy to predict.

\citet{skinner01} use this dataset to derive distances of the X-ray-emitting plasma from the O star by considering
the following relationship between the $R$ ratio, the electron density \Ne\ and the UV photoexcitation rate $\phi$ from the
$f$ to $i$ upper levels:
\begin{equation}
	R = \frac{R_0}{1 + (\phi / \phi_\mathrm{c}) + (\Ne / n_\mathrm{c})}
\label{eq:Rratio}
\end{equation}
where $n_\mathrm{c}$ is the critical density and $\phi_\mathrm{c}$ the critical photoexcitation rate
\citep*{blumenthal72}. $R_0$ is the low-density, low-UV flux limit of the $R$ ratio, i.e. 
if $\Ne \ll n_\mathrm{c}$ and $\phi \ll \phi_\mathrm{c}$, $R \rightarrow R_0$.
\citet{skinner01} argue that $R \ge \frac{2}{3}R_0$ for their \fir\ triplets (as lower values of the $R$ ratio
would be unambiguously detected), and hence $\phi/\phi_\mathrm{c} \le 0.5$. Using the results of
\citet{blumenthal72}, they derive minimum line formation radii of $r_\mathrm{min} \approx 3\Rstar(\mathrm{O})$,
$9\Rstar(\mathrm{O})$ and $30\Rstar(\mathrm{O})$ for \SiXIII, \MgXI\ and \NeIX, respectively. Their results for
\SiXIII\ and \NeIX\ are consistent with our results. Using the more recent results of \citet{porquet01}
instead of \citet{blumenthal72}, we have increased $r_\mathrm{min}$ for \SiXIII\ from $3\Rstar(\mathrm{O})$
to $5\Rstar(\mathrm{O})$. On the other hand, our analysis does not place a strong constraint on $r_\mathrm{min}$
for \NeIX, mainly because \citet{porquet01} do not quote values of the $R$ ratio for dilution factors below 0.01.

It is not surprising that \NeIX\ originates far from the O star. Near the line of centres (and hence near the O
star) the winds collide head-on, producing shock-heated plasma that is too hot for \NeIX\ to exist (tens of\MK).
This plasma moves outwards along the shock-cone, cooling radiatively. It will not cool to the temperature at which
\NeIX\ emission peaks ($\approx$4\MK) until it has travelled some distance (the hydrodynamical simulations discussed
in Section~\ref{sec:Modelling} indicate the plasma cools to $\approx$4\MK\ at a distance of 
$\approx$$10\Rstar(\mathrm{O})$ from the O star). This is a factor of 3 less than the minimum line formation
radius derived by \citet{skinner01}. However, an inspection of the models used in Section~\ref{subsec:2Tapec}
shows that there is considerable iron emission near the \NeIX\ \fir\ triplet; combined with the fact that there
are very few counts in this part of the spectrum, this means the \NeIX\ results are probably untrustworthy.

Our results for the location of the \MgXI-emitting plasma are different from those of \citet{skinner01}.
We find this plasma is located at $1.7\Rstar(\mathrm{O}) < r < 5\Rstar(\mathrm{O})$, whereas \citet{skinner01}
derive $r > 9\Rstar(\mathrm{O})$. Part of this discrepancy may come from the fact we measure different
values of the $R$ ratio for \MgXI: $1.8^{+0.9}_{-0.5}$ \citep{skinner01} versus $1.0 \pm 0.2$ (this work).
This difference may be due to differences in the calibration used, or to the fact that \citet{skinner01}
use background-subtracted HEG data only, whereas we use data from both gratings without background subtraction.
It should be noted that our measured width for the \MgXI\ \fir\ triplet is significantly larger than most
of the other widths in Table~\ref{tab:LineFitting}. In order to assess whether or not this affects our flux
measurements, we have repeated the fit with the triplet's FWHM fixed at 1200\kmps\ (the average FWHM of all the
lines). These measured fluxes are $\sim$10--20 per cent lower than our original measured values, but the
differences are not significant. Furthermore, the $R$ ratio is unaffected.

\citet{skinner01} further state that uncertainties in their flux measurements and possible blending with
\NeX\ Lyman lines means that their value of the \MgXI\ $R$ ratio could be consistent with the low-density
limit $R_0$. The \NeX\ Lyman lines from upper levels 6--10 have wavelengths of 9.362, 9.291, 9.246,
9.215 and 9.194\angstrom, respectively \citep{huenemoerder01}, whereas the wavelengths of the \MgXI\
forbidden and intercombination lines are 9.314 and 9.231\angstrom\ (\textsc{atomdb} v1.1.0). Given the
wavelengths of these \NeX\ lines, and without any information on their emissivities (these lines are not in
\textsc{aped}), it is difficult to say which of the $f$ or $i$ line will be more affected by blending, and
hence one cannot say for sure that the measured value of the $R$ ratio for \MgXI\ is probably underestimated.

If we do assume that blending with \NeX\ Lyman lines tends to lower the observed value of the $R$ ratio for \MgXI,
we can derive a location for the \MgXI-emitting plasma using the method in \citet{skinner01}. However, whereas
they used $R \ge \frac{2}{3}R_0$, given our results we can only say that $R \ge \frac{1}{3}R_0$. This implies
$\phi/\phi_\mathrm{c} \le 2.0$, and gives $r > 4.5\Rstar(\mathrm{O})$, which is consistent with the range we
derive above using the \citet{porquet01} data.

\subsection{Geometry of the wind-wind collision}
\label{subsec:Geometry}

\begin{figure}
\centering
\includegraphics[width=8cm]{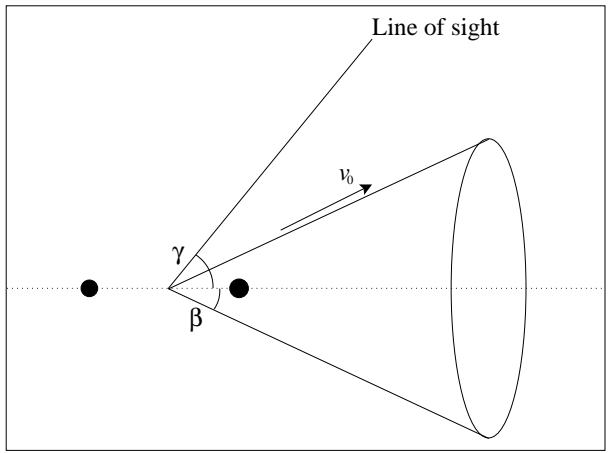}
\caption{Geometrical model of the wind-wind collision in $\gamma^2$~Vel. The solid circles represent the two
stars. The cone (with opening half-angle $\beta$) represents the wind-wind interaction region (along which X-ray-emitting
material is streaming at speed $v_0$). The viewing angle $\gamma$ is the angle between the line of sight and the
line of centres.}
\label{fig:geom_model}
\end{figure}

Some insight into the geometry of the wind-wind collision zone may be gained by assuming the X-ray line
emission comes from a cone with opening half-angle $\beta$ whose symmetry axis lies along the line of
centres with the apex pointing towards the WR star \citep{luhrs97,pollock04}. This arrangement is
illustrated in Fig.~\ref{fig:geom_model}. The angle $\gamma$
between the line of centres and the line of sight is the supplement of the angle $\theta$ in equation~(\ref{eq:theta})
and hence is given by $\cos \gamma = \cos \Psi \sin i$. We assume the emitting material is flowing along the
cone at speed $v_0$ and that there is no azimuthal velocity component [these are explicit assumptions of
\citeauthor{luhrs97}'~\citeyearpar{luhrs97} model]. The former is justified in the light of 2-D hydrodynamical
simulations of the colliding winds (such as those in Section~\ref{sec:Modelling}), which show that the velocity
vector is indeed directed along the shock cone. There is a small divergence in the velocity, but this will not
have a significant effect. The latter assumption is justified because an azimuthal velocity component in
the post-shock gas will require one to be present in the pre-shock gas, and this will not be the case in
a radially outflowing wind.

With these assumptions, the centroid shift ($\bar{v}$) and velocity range ($v_\mathrm{max} - v_\mathrm{min}$)
of an emission line are given by \citep{luhrs97,pollock04}
\begin{eqnarray}
	\bar{v}       &=& -v_0 \cos \beta \cos \gamma \\
\label{eq:shift}
	v_\mathrm{max} - v_\mathrm{min} &=& 2 \Delta v = 2 v_0 \sin \beta \sin \gamma
\label{eq:width}
\end{eqnarray}
In particular, $\bar{v}$ comes from `Term 1' in \citeauthor{luhrs97}' equation~(9), where his $\phi^\star$ equals
our $\Psi$ and his $v_\mathrm{WR}$ equals our $v_0$, while $\Delta v$ is equal to $v^\star$ in his
equation~(8) [see also his equations~(10) and (11)]. Note that \citet{luhrs97} does not use $\gamma$ in his
analysis, but instead expresses his velocities in terms of the orbital phase angle and inclination explicitly.

If we identify $\Delta v$ in equation~(\ref{eq:width}) with the FWHM, then
given the mean observed line shift and FWHM ($\bar{v} = -64 \pm 12 \kmps$, $\mathrm{FWHM} = 1240 \pm 30 \kmps$; see
Section~\ref{subsec:FitResults}) and $\gamma = 44\degr$ (see Section~\ref{subsec:2Tapec}), we obtain
$\beta = 87\degr$ and $v_0 = 1800 \kmps$. This inferred value of $v_0$ is rather large, given that the terminal
velocity of the WR star wind is 1500\kmps\ and the O star wind is likely to be colliding well below its terminal
velocity of 2300--2400\kmps. However, the inferred value of $v_0$ is sensitive to the relationship between
$(v_\mathrm{max} - v_\mathrm{min})$ and the FWHM [we have used $\mathrm{FWHM} = 0.5 (v_\mathrm{max} - v_\mathrm{min}$)
in a rather \textit{ad hoc} manner to characterize the line widths], and hence $v_0$ may in fact be somewhat
lower. This is not a problem for $\beta$ because it is very insensitive to the relationship between
$(v_\mathrm{max} - v_\mathrm{min})$ and the FWHM.

Note that the mean observed line shift is comparable to the absolute wavelength accuracy of the HETGS 
($\sim$100\kmps; \chandra\ Proposers' Observatory Guide, Section~8.2.4). This means that the measured value of $\bar{v}$
(and hence the derived values of $\beta$ and $v_0$) may not be completely trustworthy. However, we can
say that $| \bar{v} | < 100 \kmps$ (since we would unambiguously be able to detect a larger value
of $\bar{v}$). This implies that $\beta > 85\degr$. The consequences of this are discussed later.


\section{Line profile modelling}
\label{sec:Modelling}

In an attempt to better understand the line shifts and widths observed in the \chandra\ spectrum of $\gamma^2$~Vel,
we have calculated model X-ray line profiles using the model described in \citet{henley03}. This model uses
data from a hydrodynamical simulation of the wind-wind collision. The simulations are two-dimensional and assume
cylindrical symmetry about the line of centres. The stars' winds are assumed to be spherically symmetric, and
orbital effects are assumed to be negligible (this latter assumption is discussed later). The line profiles are
calculated by assuming each grid cell produces a thermally Doppler-broadened (i.e. Gaussian) line profile, the
amplitude of which depends on the density and temperature of the gas in the cell. The centre of the Gaussian
is shifted according to the line-of-sight velocity of gas. Since the simulations assume cylindrical symmetry,
each grid cell actually represents a ring of gas. The line-of-sight velocity of the gas varies as a function
of position around the ring. This is taken into account by dividing the ring into a number of sections
(typically 200), and calculating the contributions from each section (the line-of-sight velocity of a given section
depends on the 2-D velocity components in the appropriate hydro grid cell, the viewing angle $\gamma$, and the
azimuthal angle $\phi$ of the section around the ring). Absorption by the cold, unshocked winds of the stars is included
in the calculations. For each emitting ring section, the integral of the density along the line of sight is calculated
(where possible by taking the density directly from the hydro grid; off the grid the density is extrapolated
by assuming spherically symmetric winds). This is multiplied by the opacity (calculated for a solar
abundance plasma using \textsc{xstar}\footnote{http://heasarc.gsfc.nasa.gov/docs/software/lheasoft/xstar/\\xstar.html})
to give the optical depth $\tau$, and then the amplitude of the Gaussian profile produced by that ring section
is multiplied by $\mathrm{e}^{-\tau}$. The overall line profile is simply the sum of the contributions from the whole grid
(summing over $z$, $r$ and $\phi$).

The major factors which affect X-ray line profiles from colliding wind binaries are the mass-loss rates, the
wind speeds, the separation and the viewing angle (i.e. the angle between the line of sight and the line of
centres). As the orbit is well determined \citep{schmutz97,demarco99} the separation (0.92--0.94\AU) and
the viewing angle ($\gamma = 44\degr$; see Section~\ref{subsec:Geometry}) can be specified \textit{a priori}.
The appropriate set of wind parameters to use in the hydrodynamical simulation is less certain. \citet{barlow88}
measured the terminal velocity of the WR star to be 1520\kmps\ from the half-width of the forbidden 12.8\micron\
[\NeII] emission line. \citet{stlouis93} inferred a very similar value (1550\kmps) from their analysis of
phase-varying absorption in \textit{IUE} spectra. Given that the WR star wind is more powerful than the O
star wind, the wind-wind collision will be close to the O star. As the separation is
$\approx$60$\Rstar (\mathrm{WR})$, the WR star wind is very likely to be at its terminal velocity at the
wind-wind collision. Conversely, the O star wind ($v_\infty \approx 2300$--2400\kmps; \citealp*{prinja90};
\citealp{stlouis93}) is unlikely to be at its terminal velocity at the wind-wind collision, as evidenced
by the disappearance of the high-velocity blue absorption wing from ultraviolet P Cygni profiles obtained
by \textit{IUE} when the O star is behind the WR star \citep{stlouis93} and by
\citeauthor{willis95}'s~\citeyearpar{willis95} comparison of model X-ray spectra
with their observed \rosat\ spectra. Note that as the separation varies throughout the orbit, so too
will the pre-shock velocities.

As already mentioned in Section~\ref{sec:Introduction}, \citet{stevens96} derived from \asca\ data
a mass-loss rate for the WR star of $3 \times 10^{-5} \Msolpy$, a factor of three lower than
\citeauthor{barlow88}'s~\citeyearpar{barlow88} radio-determined value of $8.8 \times 10^{-5} \Msolpy$, which
\citet{stevens96} suggested may be due to the fact that radio observations tend to overestimate mass-loss
rates when the wind is clumped. It should be noted that both these analyses used larger, pre-\textit{Hipparcos} distances to
$\gamma^2$~Vel of 460\pc\ \citep{barlow88} and 450\pc\ \citep{stevens96}. Since the radio-determined mass-loss
rate scales as $D^{3/2}$, where $D$ is the distance \citep{wright75}, using the \textit{Hipparcos}
distance of 258\pc\ \citep{schaerer97} reduces \citeauthor{barlow88}'s~\citeyearpar{barlow88} value
to $3.7 \times 10^{-5} \Msolpy$. If one assumes the observed X-ray flux scales as $\Mdot^2$, the mass-loss rate
determined using \citeauthor{stevens96}'s~\citeyearpar{stevens96} method scales as $D$, and so their value reduces to
$1.7 \times 10^{-5} \Msolpy$. More recent values for the mass-loss rate of the WR star (which use the
\textit{Hipparcos} distance) include 1.08 or $3.06 \times 10^{-5} \Msolpy$ based on radio data (assuming a clumped or
smooth wind, respectively; \citealp*{nugis98}), and 0.93 or $3.3 \times 10^{-5} \Msolpy$ based on stellar
atmosphere modelling of \HeI\ and \HeII\ lines (the different values coming from different codes and different
filling factors; \citealp{demarco00}). However, as noted in Section~\ref{subsec:2Tapec}, the \textit{Hipparcos}
distance has recently been thrown into doubt by the discovery of an association of low-mass,
pre-main sequence stars in the direction of $\gamma^2$~Vel, and the distance to $\gamma^2$~Vel may be between
360 and 490\pc\	 \citep{pozzo00}.

For the O star, \citet{stlouis93} adopted a typical mass-loss rate for
a O9I star (the then accepted spectral classification) of $1.3 \times 10^{-6} \Msolpy$ \citep{howarth89}.
More recently, \citet{demarco99} have derived a much lower \MdotO\ of $1.8 \times 10^{-7} \Msolpy$ from 
hydrodynamical calculations of the radiatively driven wind.

In hydrodynamical simulations of the wind-wind collision in $\gamma^2$~Vel, one must take into account 
radiative cooling. This is because the cooling parameter $\chi$ (the ratio of the cooling timescale to
the flow timescale; \citealp*{stevens92}) is less than unity for the WR star wind. One effect of this
is that the wind-wind interaction region can become highly unstable \citep{stevens92}, leading to strong
time-variability of the calculated line profiles. These instabilities may be partially numerical in
origin, and so may lead to unphysical results. Furthermore, if the cooling is very strong, which will
be the case for a large adopted mass-loss rate or low wind speed [see equation~(8) in \citealp{stevens92}],
the shocked gas will collapse into a thin sheet and will not be properly resolved on the grid. This means
the model of \citet{henley03} cannot be used, as the profiles are calculated after the hydrodynamical
simulation and so do not take into account the energy that was radiated during the simulation.

The line profile calculations presented here are based on simulations carried out with the hydrodynamical code
\textsc{cobra} (see, e.g., \citealp{falle96,pittard03a}), which is second order accurate in space and time.
Radiative cooling is implemented in the simulations. A
small amount of numerical viscosity is used to damp the growth of a numerical instability
which appears when shocks are stationary on the grid. We adopt mass-loss rates of $\MdotWR = 1 \times 10^{-5} \Msolpy$ and
$\MdotO = 5 \times 10^{-7} \Msolpy$, and wind speeds of 1500\kmps\ for both winds. For simplicity, we
assume the winds are not accelerating, which is why we adopt a wind speed for the O star much lower than
its terminal velocity. Fig.~\ref{fig:COBRA} shows density, temperature and speeds map from the resulting simulation.
As can be seen, the cooling of the post-shock gas is resolved. Our simulations show little sign of instabilities or
post-shock mixing and turbulence. The lack of any instability is unsurprising for two reasons. Firstly, the wind speeds
are equal, so there will be very little slip along the contact discontinuity and the Kelvin-Helmholtz instability will
not arise. Secondly, the cooling time is long (especially for the O star wind, which remains almost adiabatic),
so the shocked gas does not collapse into a thin sheet and the thin-shell instability is not excited.
Even if turbulence is present in reality, in order to affect the results discussed below the turbulent velocity
would have to be a significant fraction of the bulk flow speed, which is unlikely to be the case.

\begin{figure}
\centering
\includegraphics[width=7cm]{fig13a.eps}
\includegraphics[width=7cm]{fig13b.eps}
\includegraphics[width=7cm]{fig13c.eps}
\caption{Density (top), temperature (middle) and speed (bottom) maps from a hydrodynamical simulation of $\gamma^2$~Vel.
The WR star is in the bottom left-hand corner of the grid; the O star is half-way along the bottom axis.
The adopted wind parameters are $\MdotWR = 1 \times 10^{-5} \Msolpy$, $\MdotO = 5 \times 10^{-7} \Msolpy$,
$\vWR = \vO = 1500 \kmps$.}
\label{fig:COBRA}
\end{figure}

We have calculated profiles for the \Lyalpha\ lines and He-like resonance lines of Ne, Mg, Si and S. The
line emissivity is assumed to vary just with temperature; we do not, for example, include the enhancement
of the He-like resonance line at high densities \citep{porquet01}, as it will not be important at the
densities present in the simulation results. The line emissivities were taken from \textsc{atomdb} v1.1.0,
which assumes solar abundances. To take into account the different compositions of the winds, we normalize
the emission from each wind using the product of the number density of the element in question and the
number density of electrons. The elemental number densities are calculated from the gas density $\rho$
using the elemental mass fractions. For the O star wind we assume solar abundances \citep{anders89},
while for the WR wind the mass fractions are calculated using $N$(H) = 0,
$N$(C)/$N$(He) = 0.14 \citep{morris99,schmutz99}, $N$(N)/$N$(He) = $1.0 \times 10^{-4}$ \citep{lloyd99},
$N$(C)/$N$(O) = 5 \citep{demarco00}, $N$(Ne)/$N$(He) = $3.5 \times 10^{-3}$ \citep{dessart00}
and $N$(S)/$N$(He) = $6 \times 10^{-5}$ \citep{dessart00}. [In the preceding, $N$(X) denotes the
\textit{number} abundance of element X.] The relative abundances of elements more massive
than neon are kept at their solar values \citep{anders89}. Note that we do not use elemental abundances
measured in Section~\ref{sec:Broadband}, as the different spectral models used give significantly
different abundances for some elements. However, in this case the shapes of the calculated profiles are not
very sensitive to the abundances used. The electron density \Ne\ is calculated from the gas density using
$\rho = 2 \mH \Ne / (1+X)$, where \mH\ is the mass of a hydrogen atom and $X$ is the mass fraction of hydrogen
\citep[equations 2.22 and 2.23]{bowers84}.
Since the line fitting described in Section~\ref{sec:LineFitting}
takes into account the presence of the fainter component of each Lyman line and the intercombination and
forbidden lines of the \fir\ triplets, the lines we calculate are singlets, enabling a direct comparison
with the shifts and widths measured from the HETGS spectrum.

\begin{figure*}
\centering
\includegraphics[width=8cm]{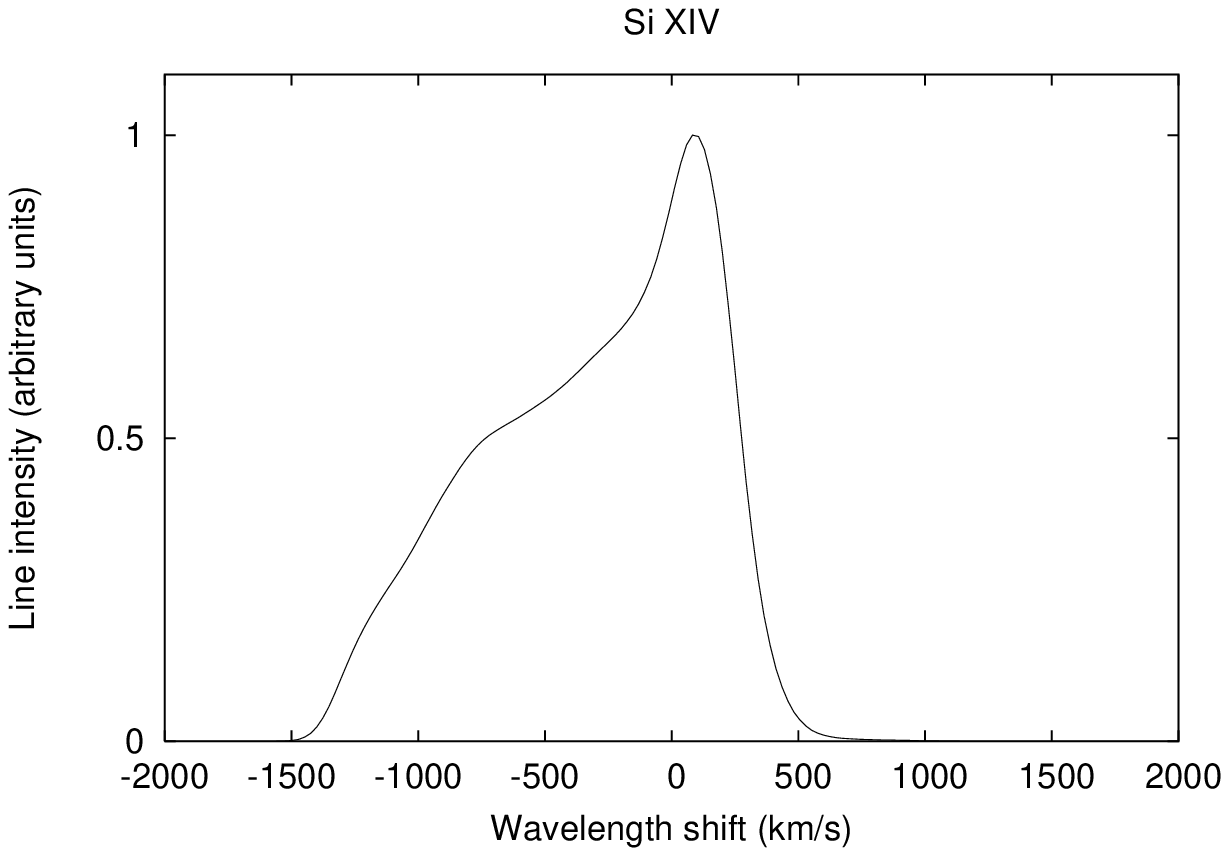}
\hspace{10mm}
\includegraphics[width=8cm]{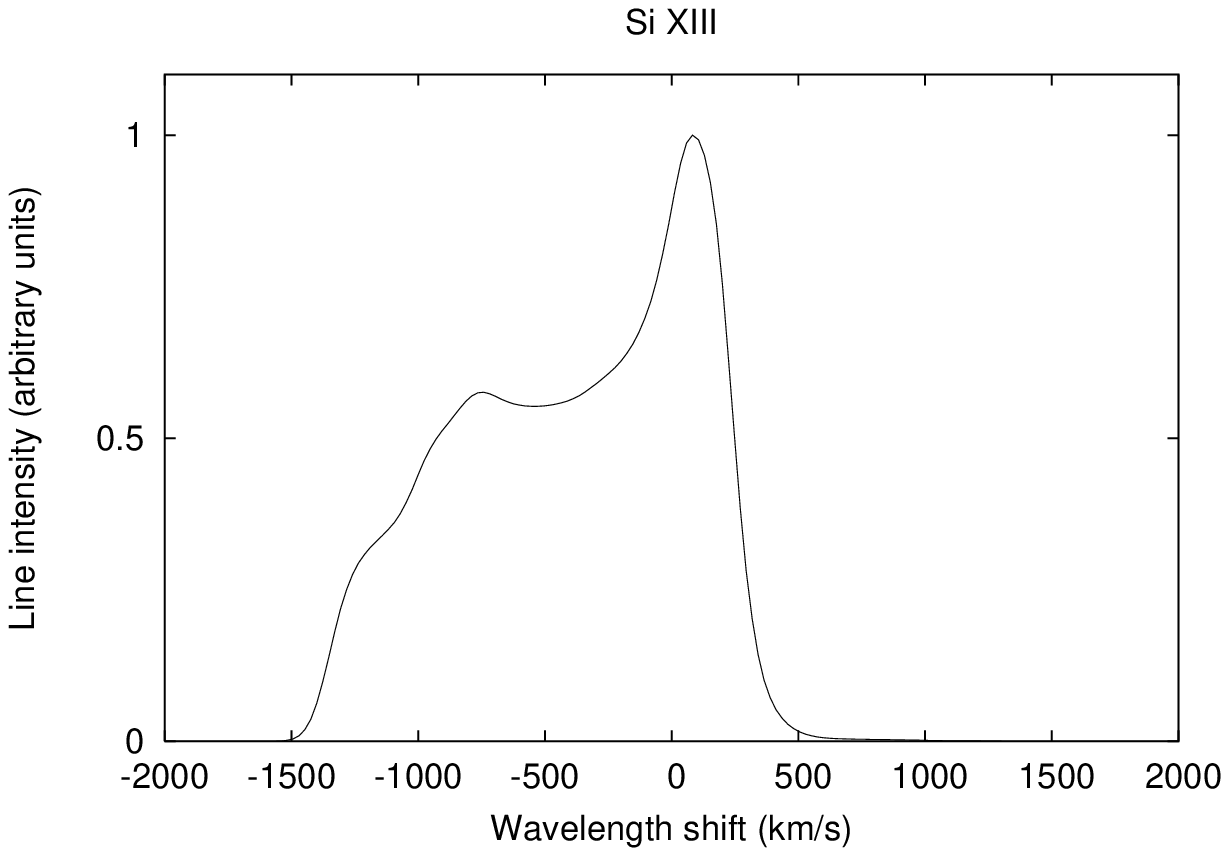}
\caption{Model \SiXIV\ \Lyalpha\ (left) and \SiXIII\ resonance (right) line profiles from $\gamma^2$~Vel
calculated for a viewing angle of $\gamma = 44 \degr$}
\label{fig:ModelProfiles}
\end{figure*}

Examples of line profiles calculated for $\gamma = 44\degr$ are shown in Fig.~\ref{fig:ModelProfiles},
which shows the \SiXIII\ resonance line and the \SiXIV\ \Lyalpha\ line. The lines are redward-skewed, 
rather than Gaussian. This skewing is intrinsic and due to the geometry of the emitting region, rather
than being due to absorption, which does not have a strong effect on these lines. The lines are broad,
each with a FWHM of $\sim$1000\kmps, in good agreement with the widths measured from the HETGS spectrum.
However, the lines are noticeably blueshifted, with centroid shifts of $\sim$$-300 \kmps$, in contrast
to the essentially unshifted lines in the observed spectrum.
The calculated line blueshift decreases as $\gamma$ increases, but given the
opening angle of the shocked region in Fig.~\ref{fig:COBRA}, increasing $\gamma$ much above $\sim$40\degr\
means the X-ray emitting plasma would be observed through the WR star wind instead of the O star wind,
and this is unlikely given the amount of absorption observed in the spectrum (the column density measured
in Section~\ref{sec:Broadband} would probably be much larger if the wind-wind collision region were
being observed through the WR star wind).

Because the line profile calculations described here are based on 2-D hydrodynamical simulations, they do not
take into account the effects of orbital motion on the shape of the wind-wind collision region. However, one
can estimate the importance of these effects by assuming the Coriolis force deflects the whole shock cone by
some angle $\xi = \tan^{-1} (v_\mathrm{orb} / v_\mathrm{wind})$ in the orbital plane, where $v_\mathrm{orb}$
and $v_\mathrm{wind}$ are the orbital and wind speeds, respectively. At the time of the \chandra\ observation,
the speed of the O star in the rest frame of the WR star was $\approx$200\kmps. If we assume
$v_\mathrm{wind} \approx 1000 \kmps$, we find $\xi \approx 11 \degr$. This means $\Psi$
[see equation~(\ref{eq:theta})] is decreased to 25\degr, and the viewing angle $\gamma$ decreases to 36\degr.
This actually increases the blueshift that the model predicts to $\sim$350\kmps\ (in this case the Coriolis
force tends to deflect the shock cone such that there is more gas travelling towards the observer, increasing
the predicted blueshift). Using $\gamma = 36\degr$ instead of 44\degr\ in Section~\ref{subsec:Geometry} still
predicts a very large shock opening half-angle ($\beta > 86 \degr$). This simple estimate of the size of the
effect of the Coriolis force does not take into account the fact that the wind-wind interaction region will
change shape, rather than simply being deflected en masse (the whole region becomes bent round into an
Archimedean spiral; e.g. \citealp{tuthill99}). Nevertheless, it appears that Coriolis effects cannot explain
the discrepancy between the observed and calculated line shifts.

The discrepancy between the observed and calculated line shifts is most likely due to the opening
half-angle of the shocked region being $\sim$40\degr\ in our hydrodynamical simulation (see Fig.~\ref{fig:COBRA}),
whereas the simple geometrical model of the emission lines in Section~\ref{subsec:Geometry} implies a
shock-cone opening half-angle of $> 85 \degr$. In a CWB with non-accelerating winds, $\beta$ is a function
only of the wind momentum ratio ($\dot{M}_\mathrm{WR} v_\mathrm{WR} / \dot{M}_\mathrm{O} v_\mathrm{O}$).
The shock opening angle in our hydrodynamical simulation
therefore depends on the adopted values of the wind parameters. However, the shock opening
half-angle inferred from our simple geometrical model ($\beta > 85\degr$) implies approximately
equal wind momenta, which is not the case for any sensible set of (constant velocity) wind
parameters. We therefore suggest that this
large implied opening angle may be evidence of sudden radiative braking
\citep*{gayley97}, in which the wind of the WR star is rapidly decelerated when it encounters the radiation
field of the O star. \citet{gayley97} suggest that the wind-wind collision in $\gamma^2$~Vel
could be significantly affected by sudden radiative braking, as they find that the wider separation
than in V444 Cygni (the main system they study) offers greater opportunities for braking.


\section{Discussion}
\label{sec:Discussion}

The broad-band X-ray spectrum of $\gamma^2$~Vel indicates that a wide range of temperatures is present
in the X-ray-emitting plasma, covering a range of $\sim$4--40\MK. The temperature distribution below
$\sim$4\MK\ is poorly contrained, because of the low number of counts above $\sim$14\angstrom. The
differential emission measure (Fig.~\ref{fig:DEM}) indicates the temperature distribution falls off
above $\sim$40\MK. This temperature range implies shock speeds of 500--1700\kmps\ assuming solar
abundances, or 300--1100\kmps\ assuming WC8 abundances. The upper values of these ranges are reasonable
given the discussion of wind speeds in the previous section. The post-shock temperature corresponding
to $v_\infty(\mathrm{O}) = 2400\kmps$ is 80\MK. The lack of any evidence for gas at such high temperatures
suggests that the O star wind is not colliding at its terminal velocity near the line of centres
[the shocked gas near the line of centres is the densest in the collision region, and so if there is
any very hot ($\sim$80\MK) gas it should make a significant contribution to the X-ray emission].
However, it should be noted that the sensitivity of the HETGS falls off at short wavelengths
($\lambda \la 2 \angstrom$) and so it is less sensitive to emission from very hot gas.
To investigate this further, we synthesized a spectrum by adding a 80\MK\ \texttt{apec} component
to our best-fitting 2$T$ \texttt{apec} model (model B in Table~\ref{tab:Broadband}). With
its emission measure [expressed in terms of the helium density
($\mathrm{EM} = \int \Ne n_\mathrm{He} \mathrm{d}V$), because of the lack of hydrogen in the WR star
wind] fixed at $2 \times 10^{53} \pcc$, this model leads to significant \FeXXV\ emission near
1.85\angstrom\ which is not observed. This implies that the emission measure of the very hot (80\MK)
gas is at least a factor of 5 smaller than those of the two \texttt{apec} components
in Table~\ref{tab:Broadband}.

The lower limits of the implied shock speeds do not imply low wind speeds \textit{per se}, but instead
that parts of the wind are being shocked obliquely as is expected from the geometry of the wind-wind
collision region. This oblique shocking also explains (at least qualitatively) the shape of the DEM in
Fig.~\ref{fig:DEM}. If all the gas were being shock-heated to $\sim$40\MK\ at the shock apex and then
allowed to cool radiatively, one would expect the DEM to have a shape that is the inverse of the cooling
function, i.e., at temperatures where the cooling function is low, there will be more gas (as it will take
longer for gas to cool through this temperature) and so a larger emission measure (and vice versa).
Therefore, one would expect the emission measure to decrease from $\sim$20\MK\ to lower temperatures
(see, e.g., fig.~3 in \citealp*{antokhin04}). Instead, the observed emission measure rises from
$\sim$20\MK\ to $\sim$8\MK. This means that gas is being fed in at lower temperatures, which is
consistent with the oblique shocking that occurs away from the line of centres.

The column density derived from the broad-band fitting ($\NH \approx 2 \times 10^{22} \pcmsq$) is consistent
with that measured by \citet{rauw00} from an \asca\ observation at the same phase
($\NH = 2.4 \times 10^{22} \pcmsq$ at $\Phi = 0.078$). The measured value indicates the wind-wind interaction
region is being observed through the wind of the O star, which is what is expected given the viewing
orientation.

As has already been stated, the broad-band spectral models used in Section~\ref{sec:Broadband} give
poor fits to the data ($\rchisq \sim 2$). Furthermore, they can at best only characterize the general
properties of the X-ray-emitting plasma. A better approach is to calculate the X-ray emission expected
from the hydrodynamics of the wind-wind collision. This may be done using numerical hydrodynamical
simulations \citep[e.g.][]{stevens96}, in a similar fashion to the method used in Section~\ref{sec:Modelling}.
An alternative is to use the semi-analytical approach devised by \citet{antokhin04}. Firstly,
the shape of the interaction region is calculated using wind momentum balance. The shocks are assumed
to be highly radiative, so the shocked material collapses into a thin layer and the two shocks and the
contact discontinuity (CD) are treated as effectively synonymous. At the shocks (which are treated as
locally planar), the wind is shock-heated to some temperature $T_\mathrm{s}$ and then cools
rapidly (effectively to $T=0$) in a thin layer between the shock and the CD; i.e., all the kinetic energy
associated with the velocity component locally perpendicular to the shock is radiated away.
The former approach works best for adiabatic shocks (the instabilities of
radiative shocks lead to substantial mixing between hot and cold material, the physically appropriate
degree of which is difficult to predict \textit{a priori}), whereas
\citeauthor{antokhin04}'s~\citeyearpar{antokhin04} method assumes fully radiative shocks. In $\gamma^2$~Vel,
the O star wind is effectively adiabatic, and while the WR star wind is radiative, it is not highly
radiative (the post-shock cooling is resolved in Fig.~\ref{fig:COBRA}). Nevertheless, a useful next step
in our understanding will be the application of both methods to the \chandra\ HETGS spectrum of
$\gamma^2$~Vel. Since velocity information is available when synthesizing spectra using either method,
it should also be possible to calculate line shapes self-consistently. Both methods may be used to generate
grids of spectra for a range of parameters (e.g. \Mdot, $v_\infty$). These can then be fit to the spectra
in order to constrain wind parameters. In order to fit to the wealth of detail in the HETGS spectrum,
the most up-to-date atomic data available must be used to synthesize the spectra. However, such an
analysis is beyond the scope of the current paper.

From the emission lines in the X-ray spectrum, we measured temperatures by comparing the fluxes of lines
from H-like and He-like ions of a given element, and by using $G$ ratios derived from He-like \fir\
triplets. For Ne and Mg, the temperatures measured by the two methods are in good agreement with
each other. However, the \SiXIV:\SiXIII\ flux ratio implies a significantly higher temperature
(12\MK) than the \SiXIII\ $G$ ratio (5\MK). As already stated, this may simply indicate that the \SiXIII\
[ionization potential (I.P.) = 2.438\kev; \citealp{dappen00}]
originates from somewhat cooler regions than the \SiXIV\ emission (I.P. = 2.673\kev). However, it could also be evidence of
non-equilibrium ionization (NEI). This could tie in with the evidence that the \MgXI\ emission (I.P. = 1.7618\kev) seems to
originate closer to the O star than the \SiXIII\ emission, as evidenced by the $R$ ratios derived from
their He-like \fir\ triplets. Furthermore, the $G$ ratios indicate that the \MgXI\ emission originates
from hotter gas ($\Te = 7 \MK$) than the \SiXIII\ emission ($\Te = 5 \MK$). This implies it originates nearer
the line of centres (and hence nearer the O star), as the temperature decreases monotonically away from the
line of centres (see Fig.~\ref{fig:COBRA}). This is counter to what is expected if collisional
ionization equilibrium holds, as in that case the \MgXI\ emission would originate in cooler gas further
from the line of centres than the \SiXIII\ emission.

One can assess whether or not NEI is important in a colliding wind binary by deriving an ionization parameter
$\psi$ analogous to the cooling parameter $\chi$ of \citet{stevens92}. In the adiabatic limit, NEI effects
will be important if the ionization equilibration timescale is larger than the flow timescale, whereas in the 
radiative limit they will be important if the ionization equilibration timescale is larger than the
cooling timescale. The ionization equilibration timescale $t_\mathrm{Ieq}$ is given by
\citep{masai94}
\begin{equation}
        t_\mathrm{Ieq} = \frac{10^{12} \pcc \s}{\Ne}
\end{equation}
where \Ne\ is the electron number density. For the flow timescale $t_\mathrm{flow}$ we shall use
\begin{equation}
        t_\mathrm{flow} = \frac{d}{c_\mathrm{s}}
\label{eq:tflow}	
\end{equation}
where $d$ is the distance from the star to the contact discontinuity and $c_\mathrm{s}$ is the post-shock sound speed
\citep{stevens92}. The cooling timescale $t_\mathrm{cool}$ is given approximately by \citep{stevens92}
\begin{equation}
	t_\mathrm{cool} = \frac{kT}{4 n \Lambda(T)}
\end{equation}
where $T$ is the post-shock temperature, $n$ is the number density of the wind at the shock and $\Lambda$ is the
cooling function ($\approx 2 \times 10^{-23} \ergps \cc$ for solar-abundance material; see fig.~10 in \citealp{stevens92}).

Note that $t_\mathrm{flow}$ in equation~(\ref{eq:tflow}) is the timescale for the post-shock gas to travel a distance $\sim$$d$
away from the line of centres, whereas the temperature in the post-shock gas may decrease away from the line of
centres such that the ionization balance for a given element is expected to go from being dominated by H-like ions to being
dominated by He-like ions in a distance less than $d$. It is therefore possible that the above flow timescale
is too long to be appropriate. Nevertheless, with this choice of $t_\mathrm{flow}$ one finds that in the adiabatic limit
\begin{equation}
        \psi_\mathrm{adiabatic} \equiv \frac{t_\mathrm{Ieq}}{t_\mathrm{flow}} = 0.02 \mu_\mathrm{e} \frac{d_{12} v_8^2}{\Mdot_{-7}}
\label{eq:psi_adiabatic}
\end{equation}
whereas in the radiative limit
\begin{equation}
	\psi_\mathrm{radiative} \equiv \frac{t_\mathrm{Ieq}}{t_\mathrm{cool}} = 0.03 \frac{\mu_\mathrm{e}}{\mu^2} \frac{1}{v_8^2}
\label{eq:psi_radiative}
\end{equation}
where $\mu_\mathrm{e}$ is the average mass \textit{per electron} in a.m.u. (1.2 for solar abundance material and
2 for WC material), $\mu$ is average mass \textit{per particle} in a.m.u. (0.61 for solar abundance
material and 1.5 for WC material), $d_{12}$ is the distance from the star to the contact discontinuity in units of
$10^{12} \cm$, $v_8^2$ is the wind speed in units of 1000\kmps, and $\Mdot_{-7}$ is the mass-loss rate in units of
$10^{-7} \Msolpy$. If $\psi \ll 1$ then collisional ionization equilibrium holds.

To determine $\psi$ we assume $\MdotWR = 3 \times 10^{-5} \Msolpy$, $\MdotO = 5 \times 10^{-7} \Msolpy$ and
$\vWR = \vO = 1500 \kmps$, and calculate $d_{12}(\mathrm{WR})$ and $d_{12}(\mathrm{O})$ by assuming ram pressure
balance at the contact discontinuity \citep{stevens92}. Using equation~(\ref{eq:psi_adiabatic}), we find
$\psi_\mathrm{WR} = 0.004$ and $\psi_\mathrm{O} = 0.02$, whereas equation~(\ref{eq:psi_radiative}) gives
$\psi_\mathrm{WR} = 0.01$ and $\psi_\mathrm{O} = 0.04$. Note that the value of $\psi_\mathrm{WR}$ in the radiative
limit is probably underestimated by a factor of a few, because $\Lambda$ is larger for WC material than solar
abundance material (see fig.~10 in \citealp{stevens92}). The winds in $\gamma^2$~Vel are neither adiabatic
nor completely radiative, and so the true values of $\psi_\mathrm{WR}$ and $\psi_\mathrm{O}$ lie between the
above values. This therefore suggests that collisional ionization equilibrium should hold for both stars' winds
in $\gamma^2$~Vel, in contrast to what we infer from the \MgXI\ and \SiXIII\ \fir\ triplets. A detailed
self-consistent calculation of the ionization balance in the wind-wind collision is needed to resolve this issue.

As well as inferring the temperatures and locations of different regions of X-ray emitting plasma from the
emission line spectrum, we have used the line shifts and widths to infer the geometry of the wind-wind
collision. The essentially unshifted lines imply a large shock-cone opening half-angle $\beta > 85 \degr$. As
stated previously, this may be evidence of sudden radiative braking \citep{gayley97}. Note that the calculation
that leads to this opening angle does not take into account absorption in the cool, unshocked wind of the O star,
which can have a profound affect on the shape of the X-ray emission lines \citep{henley03}. However, the
calculations in Section~\ref{sec:Modelling}, while they do not reproduce the observed line shifts, do 
indicate that absorption does not have a strong effect on the line profiles in $\gamma^2$~Vel.

Earlier observations of $\gamma^2$~Vel, however, do not imply such a wide shock opening half-angle. From the duration
of the period of enhanced emission in the \rosat\ lightcurve, \citet{willis95} infer a shock opening half-angle
of $\sim$25\degr. Their estimate implicitly assumes an inclination of 90\degr, but even using the inclination
we have adopted throughout this paper \citep[$i = 63\degr$;][]{demarco99} only increases the inferred half-angle
to $\sim$50\degr. Furthermore, if the shock opening half-angle is as large as 85\degr,
\citeauthor{rauw00}'s~\citeyearpar{rauw00} \asca\ observation at phase $\Phi = 0.978$ would have seen the
wind-wind collision through the O star wind, and their measured column density would have been similar
to that which they measured for their $\Phi = 0.078$ observation ($\NH = 2.4 \times 10^{22} \pcmsq$),
rather than three times larger ($\NH = 7.5 \times 10^{22} \pcmsq$).

This discrepancy between the shock opening angle inferred from the line shifts and the previous X-ray observations
of $\gamma^2$~Vel leads one to speculate that the intrinsic emission of the O star is contributing a significant
amount to the observed X-ray spectrum (the WR star is unlikely to contribute any intrinsic emission, as no single
WC star has ever been convincingly detected in X-rays; \citealp{oskinova03}).
The standard wind-shock model, with X-rays being emitted from
shocks distributed throughout the wind due to instabilities in the line-driving force \citep[e.g.][]{owocki88},
predicts broad, blueshifted, blue-skewed line profiles \citep{ignace01a,owocki01}, as the red-shifted emission from
the far side of the wind is attenuated by the wind. Such profiles have been observed from the well-studied O4f
star $\zeta$~Pup \citep{cassinelli01a}, and have been successfully fitted with a simple model of
wind absorption \citep*{kramer03a}, but in general a clear picture of X-ray line emission from early-type stars
has yet to emerge. The models for X-ray line profiles from single O stars discussed above cannot be applied
to $\gamma^2$~Vel because they rely upon spherical symmetry, which is not the case in $\gamma^2$~Vel because
of the wind-wind collision. However, if the lines were coming from the O star wind, the FWHM (1200\kmps) implies
that most of the emission originates from where the line-of-sight velocity is between $-600$ and $+600 \kmps$
(i.e. from where $|v| \la v_\infty / 4$). If one assumes a $\beta$ velocity law of the form
$v(r) = v_\infty(1 - \Rstar/r)^\beta$ with $\beta = 0.8$ \citep{stlouis93}, this implies the line emission
mostly originates less than $1.2 \Rstar(\mathrm{O})$ from the centre of the O star, which is inconsistent
with the minimum radii of formation inferred from the $R$ ratios of He-like ions. Of course, this argument
says nothing about the lines from H-like ions, but given that the widths and shifts are similar for all
ions (Fig.~\ref{fig:LineFittingResults}), it is reasonable to assume a common origin for all the lines.

The $\Lx / \Lbol$ ratio could be used to assess the importance of the O star's intrinsic emission to the total
emission observed from $\gamma^2$~Vel. \citet{pollock87} measured $\Lx / \Lbol = 0.44 \times 10^{-7}$ for $\gamma^2$~Vel
with \einstein, which is typical for lone O stars \citep[$\Lx / \Lbol \sim 10^{-7}$;][]{chlebowski89b,moffat02}. 
Model B in Table~\ref{tab:Broadband} gives an intrinsic X-ray luminosity in the \einstein\ (0.2--3.5\kev)
band of $9.6 \times 10^{32} \ergps$. Using $\Lbol(\mathrm{O}) = 2.1 \times 10^5 \Lsol$ \citep{demarco99}
gives  $\Lx / \Lbol = 12 \times 10^{-7}$, which is much higher than the \einstein\ value and which
seems to indicate that colliding wind emission dominates. Unfortunately, due to the low number of counts
at low energies, the luminosity at lower energies is very poorly constrained, and so this conclusion may
be unreliable. However, the high temperatures inferred from the spectrum (up to $\sim$40\MK) are
evidence of colliding wind emission, as the standard wind-shock model does not produce strong enough
shocks to produce such high temperatures.


\section{Summary}
\label{sec:Summary}

We have carried out a new analysis of an archived \chandra\ HETGS spectrum of $\gamma^2$~Vel. As with the
previously published analysis \citep{skinner01} we find the lines are essentially unshifted, with
a mean FWHM of $1240 \pm 40 \kmps$. The lack of strong absorption means the line-of-sight is through
the O star wind. We find a wide range of temperatures is present in the X-ray-emitting
plasma, from $\sim$4 to $\sim$40\MK, which is in good agreement with the range of temperatures expected
from a hydrodynamical simulation of the wind-wind collision. However, line profile calculations based
on such simulations imply the lines should be blueshifted by a few hundred\kmps. We suggest that
the unshifted lines may imply a wider shock opening half-angle ($> 85 \degr$) than that which is seen
in the hydrodynamical simulations. This may be evidence of sudden radiative braking. However, such a wide
shock opening half-angle seems to disagree with earlier \rosat\ and \asca\ observations of $\gamma^2$~Vel.
A full radiation hydrodynamics simulation of $\gamma^2$~Vel is needed to investigate what effect
radiative braking has on the morphology of the wind-wind collision and how the degree of this effect may
change with orbital separation \citep[see][]{pittard98b}.

Unlike \citet{skinner01}, we find evidence that the \MgXI\ emission emerges from hotter gas closer to the O star
than the \SiXIII\ emission, which may be evidence of non-equilibrium ionization. Future work will include
trying to model the ionization balance in colliding wind binaries self-consistently. This may be important
for the study of other colliding wind systems, such as WR140 \citep{pollock04}.

\citet{gayley97} point out that the eccentricity of $\gamma^2$~Vel allows the degree of braking to be studied
as a function of the orbital separation. Unfortunately, the strong absorption in the WR star wind means that
a long ($\sim$200~ks) \chandra\ pointing would be needed to obtain a high-quality spectrum 
at phases other than when the O star is in front. However, $\gamma^2$~Vel is an ideal target for
\textit{Astro-E2}, due to its greater effective area at higher energies. We have proposed a 75~ks
\textit{Astro-E2} observation of $\gamma^2$~Vel when the O star is not in front (PI: D. Henley).
This will enable us to
measure line shifts and widths at a different orbital phase from the \chandra\ observation.
From this we can investigate how the shock opening angle varies with orbital phase. By comparing this
with the shock opening angle inferred from radiation hydrodynamics simulations, we will be able to
place constraints on the coupling between the O star radiation field and the WR star wind, which will
in turn provide insights into the poorly understood physics of WR wind driving.


\section*{Acknowledgments}

DBH gratefully acknowledges funding from the School of Physics \& Astronomy.
We would like to thank Andy Pollock for useful discussions on emission line fitting,
and the referee, David Cohen, whose comments have significantly improved this paper.
This research has made use of the SIMBAD database, operated at CDS, Strasbourg, France.
This research has also made use of the \chandra\ Data Archive (CDA), part of the \chandra\ X-Ray Observatory
Science Center (CXC), which is operated for NASA by the Smithsonian Astrophysical Observatory.


\bibliography{references}


\end{document}